\documentclass[12pt, draftclsnofoot, onecolumn]{IEEEtran}
\usepackage{color}
\usepackage{graphicx}
\usepackage{epstopdf}
\usepackage{amsthm}
\usepackage{amssymb}
\usepackage{epsfig}
\usepackage{amsfonts}
\usepackage[]{algorithmic}
\usepackage[]{algorithm2e}
\usepackage{listings}
\usepackage[keeplastbox]{flushend}
\usepackage{fancybox}
\usepackage{epsfig}
\usepackage{mathtools}
\usepackage[]{algorithm2e}

\usepackage{amsmath}
\ifCLASSINFOpdf
\else
\fi
\newcommand{\Ntx}{N_{\mathrm{tx}}}
\newcommand{\Nrx}{N_{\mathrm{rx}}}
\newcommand{\Hel}{\mathbf{H}\left[ \ell \right]}
\newcommand{\tn}{\mathbf{t}\left[ n \right]}

\allowdisplaybreaks
\usepackage{graphicx}
\usepackage{amssymb}
\usepackage{xcolor}
\usepackage{subfigure}
\usepackage{cite}
\usepackage{url}
\usepackage[normalem]{ulem} 
\usepackage{etoolbox} 
\usepackage{empheq}
\usepackage{psfrag}

\usepackage{enumitem} 
\setlist[enumerate]{leftmargin=1.25em, itemsep=1ex} 
\setlist[itemize]{leftmargin=1.0em, itemsep=1ex} 

\newtoggle{long}
\toggletrue{long} 
\iftoggle{long}{
}{
}
\newcommand{\longcolor}{black} 

{\bf}{\it}

\DeclareMathAlphabet{\mathsfbf}{OT1}{cmss}{bx}{n}
\DeclareSymbolFont{bbold}{U}{bbold}{m}{n}
\DeclareSymbolFontAlphabet{\mathbbold}{bbold}
\renewcommand{\hat}{\widehat}

\renewcommand{\bar}{\overline}

\makeatletter
\newcommand{\vast}{\bBigg@{3}} 
\newcommand{\Vast}{\bBigg@{4}}
\makeatother

\newcommand{\ba}[1]{\begin{array}{#1}}
\newcommand{\ea}{\end{array}}
\newcommand{\beq}{\begin{equation}}
\newcommand{\eeq}{\end{equation}}
\newcommand{\beqar}{\begin{eqnarray}}
\newcommand{\eeqar}{\end{eqnarray}}
\newcommand{\beqars}{\begin{eqnarray*}}
\newcommand{\eeqars}{\end{eqnarray*}}

 \newcommand{\defn}{\triangleq}

 \newcommand{\of}[1]{^{(#1)}}

 \DeclareMathOperator{\E}{E}
 \DeclareMathOperator{\var}{var}

 \renewcommand{\eqref}[1]{(\ref{#1})}
 \newcommand{\figref}[1]{Fig.~\ref{#1}}
 \newcommand{\tabref}[1]{Table~\ref{tab:#1}}
 \newcommand{\secref}[1]{Section~\ref{sec:#1}} 

 \newcommand{\putTable}[3]{\begin{table}[tp]
  			    \centering
     			    #3
     			    \caption{#2}
     			    \label{tab:#1}
			  \end{table} }

 \newcommand{\giv}{\,|\,}

  \newcommand{\Br}{b} 
  \newcommand{\Nb}{N_{\mathrm{b}}} 
  \newcommand{\iB}{i} 
  
  \newcommand{\Cr}{c} 
  \newcommand{\Nc}{N_{\mathrm{c}}} 
  \newcommand{\jC}{k} 
 
  \newcommand{\Np}{N_\mathrm{p}}

  \newcommand{\zmat}[5]{{\hat{z}}_{#1}^{(#2,#3)}(#4)^{#5}}
  \newcommand{\zmatC}[5]{{z}_{#1}^{(#2,#3) #5}} 
  \newcommand{\nod}{*}

\graphicspath{{./figures/}} 

\newcommand{\remove}[1]{}

\begin{document}
\title{Message passing-based joint CFO and \\ channel estimation in millimeter wave systems\\ with one-bit ADCs}
\author{Nitin Jonathan Myers, {\it Student Member, IEEE}, and\\ Robert W. Heath Jr., {\it Fellow, IEEE}.  \thanks{ N. J. Myers (nitinjmyers@utexas.edu) and R. W. Heath Jr. (rheath@utexas.edu) are with the  Wireless Networking and Communications  Group, The University of Texas at Austin, Austin,
TX 78712 USA. This material is based upon work supported in part by the National Science
Foundation under grant numbers NSF-CCF-1527079, NSF-CNS-1702800, and by a gift from Huawei 
Technologies, Inc.}}
\maketitle
\begin{abstract}
Channel estimation at millimeter wave (mmWave) is challenging when large
antenna arrays are used. Prior work has leveraged the sparse nature of mmWave channels via compressed sensing based algorithms for channel estimation. Most of these algorithms, though, assume perfect synchronization and are vulnerable to phase errors that arise due to carrier frequency offset (CFO) and phase noise. Recently sparsity-aware, non-coherent beamforming algorithms that are robust to phase errors were proposed for narrowband phased array systems with full resolution analog-to-digital converters (ADCs). Such energy based algorithms, however, are not robust to heavy quantization at the receiver. In this paper, we develop a joint CFO and wideband channel estimation algorithm that is scalable across different mmWave architectures. Our method exploits the sparsity of mmWave MIMO channel in the angle-delay domain, in addition to  compressibility of the phase error vector. We formulate the joint estimation as a sparse bilinear optimization problem and then use message passing for recovery. We also give an efficient implementation of a generalized bilinear message passing algorithm for the joint estimation in mmWave systems with one-bit ADCs. Simulation results show that our method is able to recover the CFO and the channel compressively, even in the presence of phase noise. 
\end{abstract}

\IEEEpeerreviewmaketitle

\section{Introduction}
Millimeter wave communication introduces new challenges in the design of MIMO communication systems \cite{ranganmmwave}. For instance, large antenna arrays at the transmitter (TX) and the receiver (RX) are necessary to meet the link budget requirements \cite{mmintro}. As a result, the channel has a higher dimension compared to what is typical in lower frequency MIMO systems, and must be estimated more frequently thanks to the smaller coherence time \cite{heathwicomm}. Furthermore, cost and power consumption are major issues at the larger bandwidths that accompany mmWave, primarily due to high resolution ADCs \cite{heathoverview}. 
Typical mmWave hardwares that limit power consumption at large bandwidths introduce compression in the channel measurements. 
For example, the one-bit ADC architecture \cite{heathoverview} allows access to the output of every antenna at the expense of heavy quantization. The compression of channel measurements and the use of large antenna arrays complicate signal processing at mmWave. 
\par Compressed sensing (CS) \cite{csintro}\cite{onebit} is an efficient technique to recover sparse high-dimensional signals with few projections. As MIMO channel matrices at mmWave are sufficiently sparse when expressed in an appropriate dictionary, applying tools from CS to mmWave channel estimation can potentially  reduce the training overhead. CS-based sparse channel estimation algorithms have been proposed for various hardware architectures \cite{cschest, cshybrid ,mo2014channel}. Recent developments in approximate message passing \cite{AMP} \cite{EMGAMP} have enabled channel estimation algorithms in low resolution receivers \cite{mo2016channel}. Most CS-based channel estimation algorithms, however, assume perfect synchronization and fail in practice because of the CFO and phase noise \cite{csdanijela}.
\par CFO and phase noise are hardware impairments that corrupt the phase of the channel measurements. 
The mismatch between the carrier frequencies of the local oscillators at the TX and the RX results in CFO. Phase noise in the system arises due to short-term random fluctuations in the frequency of the oscillators. Both these non-idealities are larger at mmWave due to the high carrier frequency and ignoring them can result in significant channel estimation error \cite{nitinanalog}. Correcting for the CFO and then performing channel estimation seems like a possible solution. The disadvantage, however, is that prior to beamforming or channel estimation, mmWave systems operate at very low SNR, which can result in significant error in the CFO estimate. 
Prior work has considered joint CFO and channel estimation \cite{jointlowfreq}\cite{jointlowfreq2} in  lower frequency systems. These joint estimation algorithms, however, cannot be applied to typical mmWave systems due to differences in the hardware architectures. Furthermore, they are not designed to incorporate the sparse nature of mmWave channels. Therefore, there is a need to design either phase error robust channel estimation algorithms, or joint CFO and channel estimation algorithms that can exploit the sparsity of mmWave channel.
\par Recent work on phase error robust channel estimation is limited to narrowband systems and has focussed on specific mmWave hardware. In \cite{csdanijela},\cite{agile} and \cite{Madnonc}, phase error robust compressive beamforming algorithms were proposed for the analog beamforming architecture. In \cite{csdanijela}, phase tracking followed by phase error compensated compressive beamforming was proposed. The compensation, however, was done prior beamforming and therefore suffers from low SNR. The non-coherent algorithms in \cite{agile} and \cite{Madnonc} are not robust to heavy quantization at the receiver. These algorithms cannot be used in one-bit receivers because the energy information of the channel measurements is completely lost due to one-bit quantization. 
\par Recent joint CFO and sparse narrowband channel estimation algorithms in \cite{nitinanalog} and \cite{nitinonebit} require high computational complexity when extended to wideband systems. In \cite{nitinanalog}, we proposed a joint CFO and narrowband channel estimation algorithm using third-order tensors \cite{tensor}. We also developed a sparsity-aware joint estimation algorithm for the one-bit ADC architecture in \cite{nitinonebit}. The main idea underlying our approach in \cite{nitinonebit} was to use the lifting technique \cite{phaselift} along with message passing for the joint estimation with one-bit channel measurements. Extending the narrowband solutions in \cite{nitinanalog} or \cite{nitinonebit} to typical wideband mmWave systems would require convex optimization over millions of variables, which may be prohibitive in a practical setting. 
\par In this paper, we propose a sparse bilinear formulation of the joint CFO and wideband channel estimation problem, and solve it using message passing. We assume that all the RF chains at the TX or the RX are driven by the same reference oscillator. Hence, there is a unique CFO and a phase noise process in our MIMO system model, that corrupt the channel measurements. We also assume that there is perfect frame timing synchronization between the TX and the RX. We summarize the main contributions of our work as follows.
\begin{itemize}
  \item We formulate the joint CFO and wideband channel estimation problem as a noisy quantized sparse bilinear optimization problem. Our framework leverages the sparse nature of the wideband channel in the angle and delay domains, and also exploits the compressibility of the phase error vector in the frequency domain. 
  \item To solve the non-convex problem at hand, we use the vector variance version of the Parametric Bilinear Generalized Approximate Message Passing (PBiGAMP) algorithm \cite{PBiGAMP} and optimize it for fast joint estimation. The parameters of the sparse priors corresponding to the wideband channel and the phase error vector are learned using an Expectation Maximization (EM) algorithm.
  \item We provide insights into the design of training matrices for joint CFO and channel estimation using PBiGAMP. Specifically, we show that shifted Zadoff-Chu training proposed in \cite{mo2016channel} to accelerate message passing cannot be used for joint estimation as it results in a continuum of optimal solutions for the bilinear optimization problem. We explain the ``CFO propagation effect'' to highlight the trade-off between fast message passing and identifiability in the sparse bilinear problem.  
  \item We evaluate the performance of our joint estimation algorithm assuming a digital receiver architecture with one-bit ADCs and compare it with the hypothetical full resolution case. Simulation results show that the proposed approach is able to recover both the channel and the CFO compressively  
with IID Gaussian and IID QPSK training matrices, even in the presence of phase noise uncertainity. 
\end{itemize}
Our algorithm is advantageous over the existing sparsity-aware methods for joint estimation or phase error robust channel estimation in terms of the capability to efficiently handle frequency selective channels and scalability to other mmWave architectures.
\par \textbf{Notation}$:$ $\mathbf{A}$ is a matrix, $\mathbf{a}$ is a column vector and $a, A$ denote scalars. Using this notation $\mathbf{A}^T,\overline{\mathbf{A}}$ and $\mathbf{A}^{\ast} $ represent the transpose, conjugate and conjugate transpose of $\mathbf{A}$. The matrices $\left|\mathbf{A}\right|$ and $\left|\mathbf{A}\right|^{2}$ contain the element-wise magnitude and squared magnitude of the entries of $\mathbf{A}$. We use $\mathbf{A}^{(i)}$ and $\mathbf{A}_{(j)}$ to denote the $i^{\mathrm{th}}$ row and $j^{\mathrm{th}}$ column of $\mathbf{A}$. We use $\mathrm{diag}\left(\mathbf{a}\right)$ to denote a diagonal matrix with entries  of $\mathbf{a}$ on its diagonal. The scalar $a_m$ denotes the $m^{\mathrm{th}}$ element of $\mathbf{a}$. The symbol $\otimes$ is used to denote the kronecker product.  $\mathrm{vec}\left(\mathbf{A}\right)$ is a vector obtained by stacking all the columns of $\mathbf{A}$ and $\mathrm{vec}_m\left(\mathbf{A}\right)$ denotes the $m^{\mathrm{th}}$ element of $\mathrm{vec}\left(\mathbf{A}\right)$. We define $\mathbf{A}_{i,j}=\mathrm{vec}_i \left( \mathbf{A}_{(j)}\right)$. We use $\mathcal{I}_N$ to denote the set $\left\{ 1,2,3,..N\right\}$. The matrix $\mathbf{U}_N \in \mathbb{C}^{N \times N}$ denotes the unitary discrete Fourier transform matrix. $\mathcal{N}(\mathbf{m}, \mathbf{R})$ is the probability density function of complex Gaussian random vector with mean $\mathbf{m}$ and covariance $\mathbf{R}$. We define $\mathbf{e}^{{\ell},N} \in \mathbb{R}^{N \times 1}$ as the $N$ dimensional canonical basis vector with its $\ell^{\mathrm{th}}$ coordinate as 1.

\section{System and Channel Models} \label{sec:systemmodel}
In this section, we describe the underlying hardware architecture, CFO and phase noise model, and the wideband mmWave channel model used for our simulations. In particular, we focus on the digital receiver architecture with one-bit ADCs, to highlight the differences with the existing non-coherent algorithms. Nevertheless, our algorithm can be extended to other mmWave architectures  as the underlying joint estimation problem is bilinear in nature.
\subsection{System Model}
\begin{figure}[h]
\centering
\includegraphics[width=0.75 \textwidth]{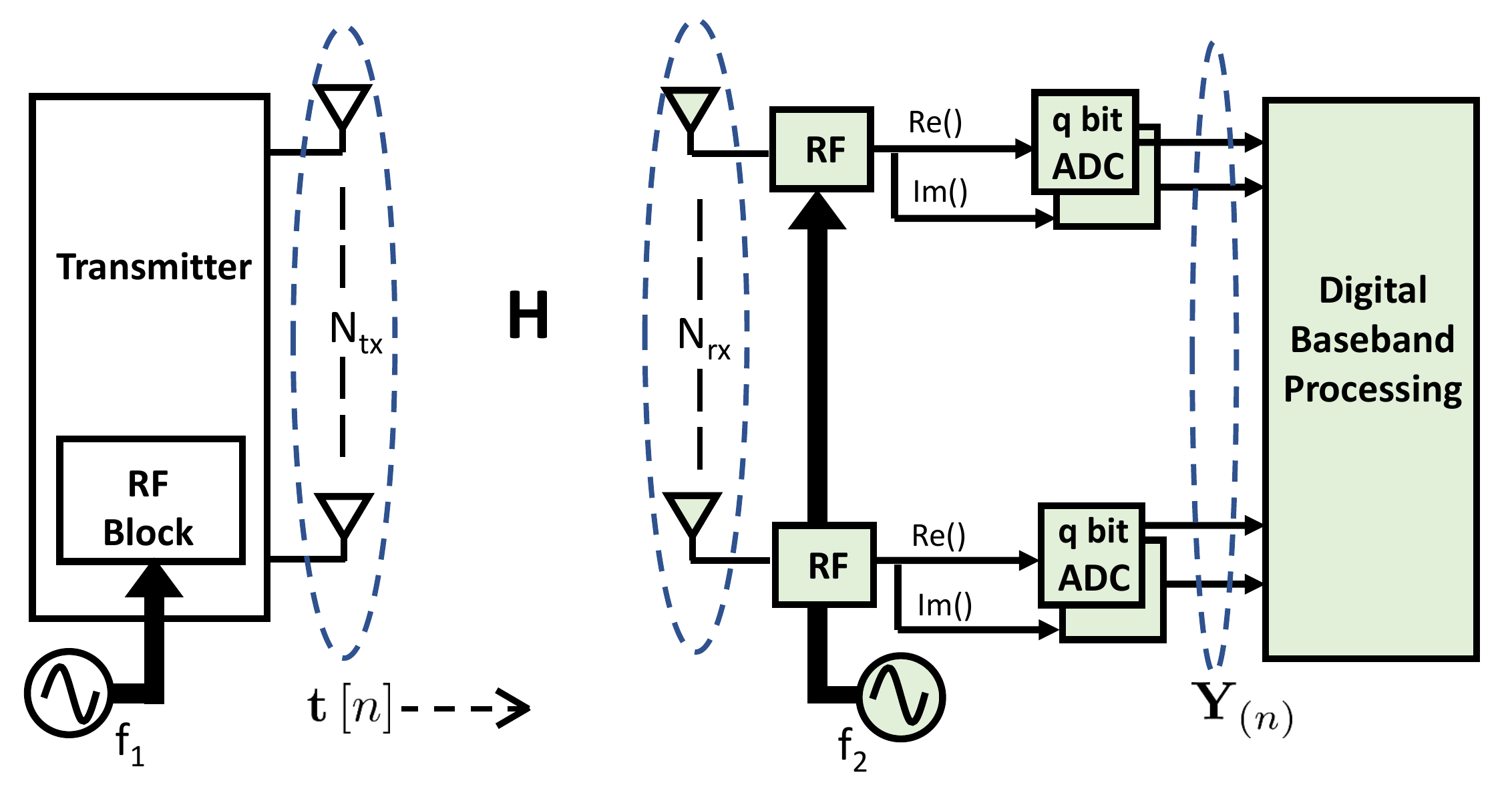}
  \caption{A MIMO system with local oscillators operating at $f_1$ and $f_2$, and $q$-bit ADCs at the receiver. Each antenna is associated with an RF chain and a pair of $q$-bit ADCs. In this work, we consider the extreme cases of $q=1$ and $q=\infty$.}
  \label{fig:architect}
\end{figure}
\par We consider a MIMO system with uniform linear array of $\Ntx$ antennas at the TX and $\Nrx$ antennas at the RX, as shown in \figref{fig:architect}. We use linear arrays for a concise representation of the simplifications involved in PBiGAMP; our framework can be extended to other array geometries using appropriate array response vectors in the formulation. We do not impose constraints on the number of RF chains or the resolution of the digital-to-analog converters (DACs) at the TX. The resolution of the $\Nrx$ ADCs at the RX, however, is assumed to be limited. The baseband signal at the TX is upconverted to the mmWave band, using a local oscillator at a carrier frequency $f_1$. The transmitted RF signal propagates through the wireless channel and is downconverted at the RX using a carrier frequency $f_2$, that slightly differs from $f_1$. Although the MIMO system can have multiple RF chains, we assume that all the RF chains at a given end are driven by the same reference oscillator. Even if the RF chains at a given end were driven by different oscillators, achieving carrier synchronization locally is feasible \cite{oscill}. After downconversion at the RX, the output at each antenna is sampled using a pair of $q$ bit ADCs, one each for the in-phase and the quadrature phase components. We use $\mathcal{Q}_q \left( . \right)$ to represent the $q$ bit quantization function corresponding to the ADCs. In this work, we consider the extreme case of $q=1$ and provide a performance comparison relative to $q= \infty$. The quantization functions for the two cases are $\mathcal{Q}_1\left(\mathbf{x}\right)=\mathrm{sign}\left(\mathrm{real}\left(\mathbf{x}\right)\right)+j\,\mathrm{sign}\left(\mathrm{imag}\left(\mathbf{x}\right)\right)$ and $\mathcal{Q}_{\infty}\left(\mathbf{x}\right)=\mathbf{x}$. Note that the functions $\mathrm{sign}\left( \cdot \right), \mathrm{real}\left( \cdot \right)$ and $\mathrm{imag}\left( \cdot \right)$ are applied element-wise on the vector. 
\par The impact of CFO on channel estimation algorithms is more significant in one-bit receivers than the full resolution ones. The mismatch in the carrier frequencies, i.e., $\left| f_2-f_1 \right|$ is typically in the order of several parts per millions (ppms) of $f_1$ or $f_2$. Due to the high carrier frequencies at mmWave, even such small differences can significantly perturb the channel estimate when ignored \cite{csdanijela}.
For a symbol duration of $T$ seconds, we define the digital domain CFO as $\epsilon= 2\pi \left(f_1-f_2 \right) T$. CFO results in unknown phase errors in the received samples that linearly increase with time.  
Hence, the impact of CFO on standard channel estimation algorithms is determined by the length of training. 
As one-bit receivers relatively need a longer training for channel estimation when compared to the full resolution ones, 
channel estimation algorithms that ignore phase errors are more vulnerable to the CFO in one-bit systems than the full resolution ones.
\par Phase noise in wireless systems arises due to jitter in the frequency of the oscillators. For a phase noise variance of $\beta^2_{\mathrm{tx}}$ at the TX and $\beta^2_{\mathrm{rx}}$ at the RX, the phase noise variance in the received samples can be approximated as $\beta^2= \beta^2_{\mathrm{rx}}+\beta^2_{\mathrm{tx}}$. The approximation is valid when the $3\, \mathrm{dB}$ bandwidth of the phase noise power spectral density is significantly smaller than the channel coherence bandwidth \cite{phasenoisesum}. Let $\phi_k$ denote the phase error introduced   in the $k^{\mathrm{th}}$ received sample, due to phase noise at the TX and the RX. As is common in prior work, we model the phase errors using a Wiener process \cite{lorentzian} in which the increments, i.e., $\phi_k -\phi_{k-1}$, are IID Gaussian random variables with zero mean and variance $\beta^2$. As $\beta^2$ is proportional to $f^2_1$ \cite{csdanijela}, phase noise is higher at mmWave carrier frequencies for a given quality of oscillator.
\par Now, we describe the received signal model in the digital receiver architecture. Let $\tn \in \mathbb{C}^{\Ntx \times 1}$ be the $n^{\mathrm{th}}$ transmit symbol satisfying the power constraint $\mathbb{E}\left[ \mathbf{t}^{\ast} \left[n\right] \tn \right]= P$. The discrete time baseband representation of the MIMO channel is assumed to be limited to $L$ taps. Let $\Hel \in \mathbb{C}^{\Nrx \times \Ntx}$ be the $\ell^{th}$ tap of the equivalent baseband channel, where $\ell \in \left\{ 0,\,1,\,2,...,\,L-1\right\} $. We assume that perfect frame timing synchronization can be achieved using the control channel. Our assumption can be justified in situations where the mmWave system co-exists with a lower frequency system \cite{anum} that can perform timing synchronization. Frequency synchronization, however, may not be achieved due to different offsets and phase noise processes for each of these systems. With the timing synchronization assumption, the sampled baseband vector in the $n^{\mathrm{th}}$ symbol duration can be given by 
\begin{equation}
\mathbf{Y}_{\left(n\right)}=\mathcal{Q}_{q}\left(e^{j\left( \epsilon n +\phi_n \right)}\sum_{\ell=0}^{L-1}\mathbf{H}\left[\ell\right]\mathbf{t}\left[n-\ell\right]+\mathbf{V}_{\left(n\right)}\right),
\label{basicmodel}
\end{equation}
where $\mathbf{V}_{\left(n\right)} \sim \mathcal{N}( \mathbf{0}, \sigma^2 \mathbf{I}_{\mathrm{\Nrx}})$ is additive white Gaussian noise. In this work, we develop an algorithm to estimate $\epsilon$ and $\left\{ \mathbf{H}\left[\ell\right]\right\} _{\ell=0}^{L-1}$ from the series of observations $\mathbf{Y}_{\left(n\right)}$. Our joint estimation algorithm can be extended to any $q$-bit ADC architecture by defining appropriate output likelihood functions in message passing.    
\subsection{Channel Model}
We consider a clustered channel model for the frequency selective mmWave MIMO channel. The channel consists of $N_{\mathrm{cs}}$ clusters with $M_n$ rays in the $n^{th}$ cluster.  Let $\gamma_{n,m}$, $\tau_{n,m}$, $\theta_{r,n,m}$ and $\theta_{t,n,m}$ denote the complex gain, delay, angle-of-arrival (AoA) and angle-of-departure (AoD) of the $m^{\mathrm{th}}$ ray in the $n^{\mathrm{th}}$ cluster. We assume that the transmitted signal is bandlimited to $1/T \, \mathrm{Hz}$. With $\omega_{r,n,m}=\pi\,\mathrm{sin}\, \theta_{r,n,m}$, $\omega_{t,n,m}=\pi \, \mathrm{sin}\, \theta_{t,n,m}$ and the Vandermonde vector
\begin{equation}
\mathbf{a}_{_N}\left(\Delta\right)=\left[1\,, e^{j\Delta}\,, e^{j2\Delta}\,, \cdots\,, e^{j(N-1)\Delta}\right]^{T},
\end{equation}
the $\ell ^{\mathrm{th}}$ tap of the wideband MIMO channel for a half wavelength spaced uniform linear array is given by
\begin{equation}
\mathbf{H}\left[\ell\right]=\sum_{n=1}^{N_{\mathrm{cs}}}\sum_{m=1}^{M_n}\gamma_{n,m}\mathbf{a}_{_{N_{\mathrm{rx}}}}\left(\omega_{r,n,m}\right)\mathbf{a}_{_{N_{\mathrm{tx}}}}^{\ast}\left(\omega_{t,n,m}\right) \mathrm{sinc} \left( \ell - \frac{\tau_{n,m}}{T}\right).
\label{wbchannel}
\end{equation}
The wideband channel can be represented using $\Nrx \Ntx L$ complex entries, and the matrix in \eqref{wbchannel} is large in typical mmWave systems. The channel impulse response in \eqref{wbchannel} is represented using a linear combination of bandlimited $\mathrm{sinc}(\cdot)$ functions. Notice that each of these $\mathrm{sinc}(\cdot)$ functions is delayed by the normalized delay spread, i.e, ${\tau_{n,m}}/{T}$ and evaluated at periodic time instants to obtain the discrete time representation in \eqref{wbchannel}. Other filtering functions could also be used to incorporate the effect of pulse shaping at the TX or filtering at the RX \cite{kiranchannel}.    
\par The mmWave MIMO channel is aproximately sparse in an appropriate dictionary due to the propagation characteristics of the environment at mmWave frequencies. Compared to the lower frequency channels, mmWave channels comprise of fewer clusters \cite{heathoverview}. Each of the channel taps $\mathbf{H}\left[\ell \right]$, is approximately sparse in the spatial Fourier basis at mmWave \cite{mo2016channel}. Furthermore, the channel is approximately sparse along the time dimension as the delays of the propagation rays are heavily clustered within the delay spread. As the delays $\tau_{n,m}$ may not necessarily be an integer multiple of $T$, there is a leakage effect along the time dimension. Let $\mathbf{C}\left[ \ell \right] \in \mathbb{C}^{\Nrx \times \Ntx}$ be the 2-D Fourier transform of $\mathbf{H}\left[ \ell \right]$, such that   
\begin{equation}
\mathbf{H}\left[ \ell \right]= \mathbf{U}_{\Nrx} \mathbf{C}\left[\ell \right] \mathbf{U}^{\ast}_{\Ntx}, \qquad \forall \ell \in \left\{ 0,\,1,\,2,...,\,L-1\right\}.
\label{mimoangledom}
\end{equation}   
The approximate sparsity of the mmWave MIMO channel along the angle and delay domains  \cite{beamsparse} is directly reflected in the matrices $\left\{ \mathbf{C}\left[\ell\right]\right\} _{\ell=0}^{L-1}$. A higher resolution dictionary can be used for the angle and delay dimensions to increase sparsity of the mmWave channel at the expense of higher dimensionality and higher frame coherence \cite{framecoh}. Our GAMP based approach will be robust to leakage effects that arise due to approximate sparsity.

\section{Demystifying joint CFO and channel estimation at mmWave}
In this section, we propose a sparse bilinear formulation for the joint estimation problem. We also identify existing techniques to solve the problem and describe their limitations in terms of scalability to other mmWave architectures and computational complexity. 
\subsection{Bilinear formulation}
\par We derive a compact form for the received signal model in \eqref{basicmodel} for a SC-FDE system \cite{heathwicomm}. Let $\mathbf{T}\in \mathbb{C}^{\Ntx \times \Np }$ be a training block of length $\Np$ such that $\mathbf{T}_{(k)}=\mathbf{t}\left[k\right]$. The TX transmits a training sequence with a cyclic prefix of length $L-1$, i.e., $\left[\mathbf{T}_{\left(\Np-L+2\right)},\mathbf{T}_{\left(\Np-L+3\right)},\,...,\mathbf{T}_{\left(\Np\right)},\mathbf{T}\right]$. The cyclic prefix padded transmission gets convolved with the frequency selective MIMO channel before sampling at the receiver. As usual, the first $L-1$ samples of the received block that experience interference from the previous transmit block are discarded \cite{heathwicomm}. Let $\mathbf{Y} \in \mathbb{C}^{\Nrx \times \Np}$ be the received block obtained after discarding the first $L-1$ received vectors. We define an $\ell$ circulant delay matrix $\mathbf{J}_{\ell} \in \mathbb{C}^{\Np \times \Np}$, such that its first column is the canonical basis vector $\mathbf{e}^{{1+\ell},\Np} $. We define a vector ${\mathbf{d}}\left(\epsilon, \beta\right) \in \mathbb{C}^{\Np \times 1}$ such that its $n^{\mathrm{th}}$ entry has the phase error corresponding to the received vector $\mathbf{Y}_{(n)}$, i.e., $d_n\left(\epsilon, \beta\right)=e^{j\left( \epsilon n +\phi_n \right)}$.  Note that $\beta$ is the standard deviation of the incremental phase errors in the Wiener phase noise process. The received samples corresponding to the transmit block defined by $\mathbf{T}$ can be expressed using \eqref{basicmodel} as   
\begin{equation}
\mathbf{Y}=\mathcal{Q}_{q}\left(\sum_{\ell=0}^{L-1} \Hel \mathbf{TJ}_{\ell}\mathrm{diag}\left({\mathbf{d}}\left(\epsilon, \beta\right)\right)+\mathbf{V}\right).
\label{blockmodel}
\end{equation} 
The phase errors in \eqref{blockmodel} are invariant along any column of the unquantized received block as there is a unique CFO and a phase noise process in the system. The representation in \eqref{blockmodel} can be further simplified to capture the structure in the phase errors and the channel. 
\par We exploit the structure in the joint estimation problem using sparsity of the channel and the phase error vector in appropriate dictionaries. A compact representation of \eqref{blockmodel} can be obtained by following the same steps in \cite{mo2016channel}, except for the diagonal matrix containing the phase errors. We use $\mathbf{b}$ to denote the DFT of the phase error vector $\mathbf{d}\left( \epsilon, \beta \right)$. With $\mathbf{Z}$ used to denote the noiseless unquantized version of $\mathbf{Y}$ in \eqref{blockmodel} such that $\mathbf{Y}= \mathcal{Q}_q \left( \mathbf{Z} + \mathbf{V} \right)$, we have 
\begin{align}
\nonumber
\mathbf{Z} & = \sum_{\ell=0}^{L-1} \Hel \mathbf{TJ}_{\ell}\,\mathrm{diag}\left({\mathbf{d}}\left(\epsilon, \beta\right)\right)\\
\nonumber
& = \sum_{\ell=0}^{L-1} \mathbf{U}_{\Nrx} \mathbf{C} \left[ \ell \right] \mathbf{U}^{\ast}_{\Ntx} \mathbf{TJ}_{\ell}\,\mathrm{diag}\left(\mathbf{U}_{\Np}^{\ast}\mathbf{b}\right)\\
&= \mathbf{U}_{\Nrx}\underbrace{\left[\mathbf{C}\left[0\right]\,\mathbf{\,C}\left[1\right]\,\,\mathbf{C}\left[2\right]\,\,.\,.\,.\,\,\mathbf{C}\left[L-1\right]\right]}_{\overset{\Delta}{=}\mathbf{C}}\underbrace{\left[\begin{array}{c}
\mathbf{U}_{\Ntx}^{\ast}\mathbf{TJ}_{0}\\
\mathbf{U}_{\Ntx}^{\ast}\mathbf{TJ}_{1}\\
\vdots\\
\mathbf{U}_{\Ntx}^{\ast}\mathbf{TJ}_{L-1}
\end{array}\right]}_{\overset{\Delta}{=}\mathbf{F}}\mathrm{diag}\left(\mathbf{U}_{\Np}^{\ast}\mathbf{b}\right).
\label{z_pre_equation}
\end{align}
The matrix $\mathbf{C} \in \mathbb{C}^{\Nrx \times \Ntx L}$ is just a concatenation of the angle domain representation in \eqref{mimoangledom} corresponding to each tap of the MIMO channel and is approximately sparse. Furthermore, the vector $\mathbf{b}$, i.e., the DFT of $\mathbf{d}\left( \epsilon, \beta\right)$ can be considered to be approximately sparse. The sparse representation of the phase error vector is valid in practice as the spread of the oscillator's spectrum about the center frequency is relatively small compared to the bandwidth of the signal.
\par Now, we derive a sparse bilinear formulation for the joint estimation problem. Using \eqref{z_pre_equation}, the quantized received block $\mathbf{Y}$ in \eqref{blockmodel} can be expressed as 
\begin{equation}
\mathbf{Y}= \mathcal{Q}_q \left( \mathbf{U_{\Nrx}C}\mathbf{F} \mathrm{diag}\left( \mathbf{U}^{\ast}_{\Np} \mathbf{b} \right) +\mathbf{V} \right).
\label{block_model_bilin}
\end{equation}
We define the $\Np \Nrx \times 1$ vectors $\mathbf{y}=\mathrm{vec}\left(\mathbf{Y}\right)$, $\mathbf{z}=\mathrm{vec}\left(\mathbf{Z}\right)$ and $\mathbf{v}=\mathrm{vec}\left(\mathbf{V}\right)$, and the $\Nrx \Ntx L \times 1$ vector $\mathbf{c}=\mathrm{vec}\left(\mathbf{C}\right)$. In vector notation, \eqref{z_pre_equation} can be written as  
\begin{equation}
\mathbf{z}=\mathrm{diag}\left( \mathbf{U}^{\ast}_{\Np} \mathbf{b} \otimes\mathbf{a}_{\Nrx}(0)\right)\mathrm{vec}\left(\mathbf{U}_{\Nrx}\mathbf{CF}\right).
\label{z_prelim}
\end{equation}
Notice that $\mathbf{a}_{\Nrx}(0)$ is just the all ones vector in $\Nrx$ dimension. Using the property $\mathrm{vec}\left(\mathbf{PQR}\right)=\left(\mathbf{R}^{T}\otimes\mathbf{P}\right)\mathrm{vec}\left(\mathbf{Q}\right)$, the received vector $\mathbf{y}$ is expressed as 
\begin{equation}
\mathbf{y}=\mathcal{Q}_{q}\left(\mathrm{diag}\left(\mathbf{U}_{\Np}^{\ast}\mathbf{b}\otimes\mathbf{a}_{\Nrx}(0)\right)\left(\mathbf{F}^{T}\otimes\mathbf{U}_{\Nrx}\right)\mathbf{c}+\mathbf{v}\right).
\label{y_prelim}
\end{equation}
We define the matrices $\mathbf{G}= \mathbf{U}^{\ast}_{\Np} \otimes \mathbf{a}_{\Nrx} \left( 0 \right)$ and $\mathbf{A}=\left( \mathbf{F}\otimes \mathbf{U}_{\Nrx} \right)^T$ to rewrite  \eqref{y_prelim} as 
\begin{equation}
\mathbf{y}=\mathcal{Q}_{q}\left(\mathrm{diag}\left(\mathbf{Gb}\right)\mathbf{Ac}+\mathbf{v}\right).
\label{bilin}
\end{equation}
Estimating the phase errors and the channel is equivalent to estimating $\mathbf{b}$ and $\mathbf{c}$ from $\mathbf{y}$ in \eqref{bilin}. The joint estimation problem in \eqref{bilin} can be observed to be a noisy quantized bilinear problem in $\mathbf{b}$ and $\mathbf{c}$, subject to the sparsity of $\mathbf{b}$ and $\mathbf{c}$.
\subsection{Limitations of existing techniques} 
\subsubsection{CFO robust methods}
Existing sparse channel estimation methods that are robust to CFO discard the phase of the channel measurements. A hashing technique based non-coherent beam alignment algorithm was proposed in \cite{agile} for analog beamforming systems. This method, however, assumes fine control over the phase shifters, which is not necessarily the case with mmWave systems. The received signal strength (RSS) based method in \cite{Madnonc} accounts for the limited phase  control and uses pseudo-random phase shifts for compressive beam-training. The solutions in \cite{agile} and \cite{Madnonc} assume a narrowband mmWave system and perform beam-alignment with just the magnitude of the channel measurements. These methods, however, cannot be used in one-bit receivers as energy detection with undithered one-bit ADCs is not feasible unless additional circuit components are used. For example, $\mathcal{Q}_1 \left(\mathbf{r} \right)$ and $\mathcal{Q}_1 \left( \alpha \, \mathbf{r} \right)$ are the same for any $\alpha > 0$. As the only information provided by undithered one-bit ADCs is phase quantized to $4$ levels, discarding it due to phase errors leaves no information.   
\subsubsection{Joint estimation using lifting} \label{sec:lifting}
Lifting \cite{phaselift}\cite{biconvex} is a convex relaxation technique that transforms a bilinear problem to a higher dimensional one and then recovers the original vectors by solving the higher dimensional problem. We describe the lifting technique applied to the joint estimation problem in \eqref{bilin}. With $M$ denoting the number of entries in $\mathbf{y}$ or $\mathbf{z}$, i.e., $M =\Nrx \Np$, the $m^{\mathrm{th}}$ entry of $\mathbf{z}$ can be given as
\begin{align}
z_{m} & = \mathrm{vec}_m \left( \mathrm{diag} \left( \mathbf{Gb}\right)\mathbf{Ac}\right) 
\label{zdepend}\\
&=\mathbf{G}^{\left(m\right)}\mathbf{b}\mathbf{A}^{\left(m\right)}\mathbf{c}\\
&=\mathbf{G}^{\left(m\right)}\mathbf{b}\mathbf{c}^{T}\left(\mathbf{A}^{\left(m\right)}\right)^{T}\\
&=\left(\mathbf{A}^{\left(m\right)}\otimes\mathbf{G}^{\left(m\right)}\right)\mathrm{vec}\left(\mathbf{b}\mathbf{c}^{T}\right), \qquad \qquad \forall m  \in \mathcal{I}_M. 
\end{align}
We define a lifted variable $\mathbf{x} = \mathrm{vec} \left(\mathbf{bc}^T\right)$ and a measurement matrix $\mathbf{\Phi} \in \mathbb{C}^{M \times \Np \Nrx \Ntx L}$, such that $\mathbf{\Phi}^{\left( m \right)}=\mathbf{A}^{(m)}\otimes\mathbf{G}^{(m)}$. Hence, the quantized measurements in \eqref{bilin} can be expressed as
\begin{equation}
\mathbf{y}=\mathcal{Q}_q \left( \mathbf{\Phi x}+\mathbf{n} \right).
\label{lifted_eqn}
\end{equation}
The lifted vector $\mathbf{x}$ in \eqref{lifted_eqn} is sparse as it is just an outer product \cite{tensor} of the sparse vectors $\mathbf{b}$ and $\mathbf{c}$. Several CS-based algorithms \cite{onebit}\cite{ranganGAMP} can be used to recover $\mathbf{x}$ from the possibly under-determined noisy quantized system in \eqref{lifted_eqn}. Using the SVD of the higher dimensional matrix estimate, the vectors in the bilinear problem can be estimated upto a scale factor.
\par Lifting followed by the SVD was applied to joint CFO and narrowband channel estimation for one-bit receivers in our previous work \cite{nitinonebit}. The main issue in extending our method in \cite{nitinonebit} to wideband systems arises due to the large dimensionality of the lifted problem. For instance, the dimension of $\mathbf{x}$ to perform joint CFO and channel estimation in wideband systems would be $\Nrx\Ntx\Np L$. Using lifting necessarily implies solving for millions of variables for typical wideband mmWave systems, due to the large number of antennas and the need for additional pilots to compensate for the heavy quantization in low resolution systems. 
\par The limitations of the existing phase error robust and joint estimation solutions in terms of architectural scalability and computational complexity, motivate the need to develop new low complexity joint estimation algorithms that can be applied to wideband systems and low resolution receivers. 
\section{Message passing based joint CFO and channel estimation}
In this section, we give a brief introduction to PBiGAMP \cite{PBiGAMP} and discuss its application to the joint  estimation in \eqref{bilin}. We exploit the inherent structure in our problem to derive a low complexity and memory efficient implementation of PBiGAMP for the joint estimation. Furthermore, we explain the CFO propagation effect induced by special training matrices that prevents further reduction in the computational complexity.   
\subsection{Introduction to PBiGAMP}
\par The joint estimation problem in \eqref{bilin} can be solved using PBiGAMP \cite{PBiGAMP} by considering $\mathbf{b},\mathbf{c}$, $\mathbf{z}$ and $\mathbf{y}$ of \eqref{z_prelim} and \eqref{bilin} as realizations of random vectors, say $\boldsymbol{\mathsf{b}},\boldsymbol{\mathsf{c}},\boldsymbol{\mathsf{z}}$ and $\boldsymbol{\mathsf{y}}$. Let $\mathsf{b}_i$, $\mathsf{c}_k$, $\mathsf{z}_m$ and $\mathsf{y}_m$ be the elements of these random vectors.
In general, deriving the closed form Minimum
Mean-Squared Error (MMSE) estimates \cite{SMkay} of $\mathbf{b}$ and $\mathbf{c}$ is difficult as it requires marginalizing the joint PDF of $\boldsymbol{\mathsf{b}}$ and $\boldsymbol{\mathsf{c}}$ conditioned on $\boldsymbol{\mathsf{y}}=\mathbf{y}$. Using ideas from message passing, PBiGAMP can obtain the MMSE estimates of both the vectors in the bilinear problem.  
\par
\begin{figure}[h]
\centering
\includegraphics[trim=4cm 2cm 6cm 1cm,clip=true,width=0.8 \textwidth]{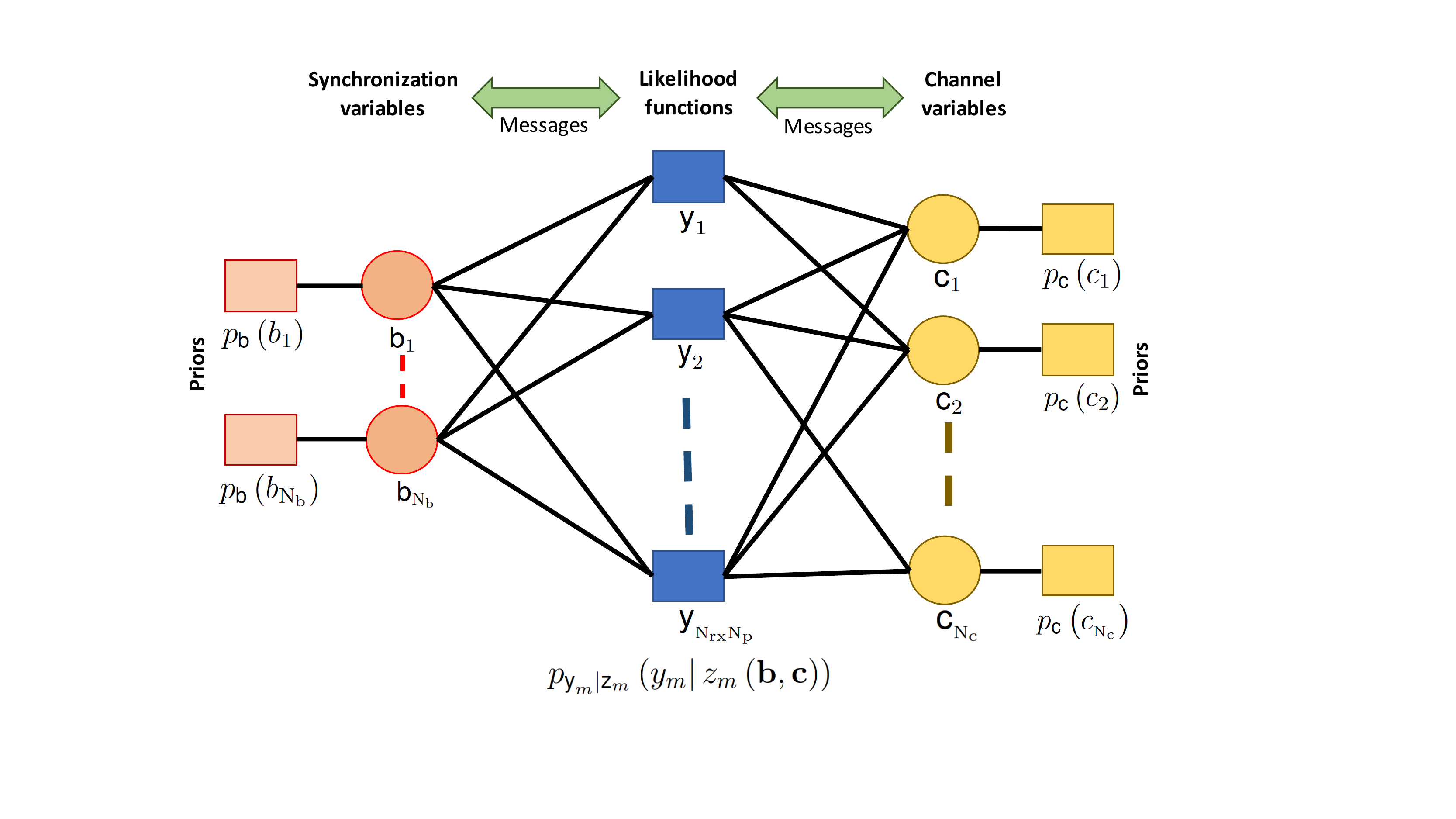}
  \caption{The factor graph of bilinear message passing for joint CFO and channel estimation. The rectangular nodes, called as factors, contain the likelihood functions corresponding to the received samples or the sparse priors. Messages are sent between the factor nodes and the variable nodes until the marginal probability distributions of the variables converge.}
  \label{fig:factor}
\end{figure}

 We explain message passing using the factor graph \cite{factorgraph} in \figref{fig:factor}, that shows the dependency between $\boldsymbol{\mathsf{y}}$, the random variables ($\boldsymbol{\mathsf{b}},\boldsymbol{\mathsf{c}}$) and their prior distributions. The circular nodes in the factor graph are called as variable nodes as they represent the random variables. The rectangular nodes in the factor graph are called as factor nodes and they contain the prior distribution of a random variable or the likelihood function associated with an observation. The  messages in message passing are essentially probability distributions, also called as beliefs. The idea underlying message passing is to perform belief flows iteratively between the factors and the variables until all the variable nodes reach a consensus on their marginal probability distributions. As the factor graph for joint estimation is strongly connected, standard message passing can be computationally  intractable. 
\par Using ideas from Approximate Message Passing \cite{AMP}, PBiGAMP simplifies the messages by assuming a large number of variable nodes. Simulation results in \cite{PBiGAMP} that show that PBiGAMP outperforms lifting techniques in \secref{lifting}, for IID Gaussian measurement matrices, motivate applying it to our problem. Furthermore, PBiGAMP performs optimization over the same number of variables in the problem, unlike lifting \cite{phaselift} that solves the problem in a higher dimensional space. For the joint estimation problem in typical wideband mmWave systems, PBiGAMP is memory efficient over lifting by several orders of magnitude.  
\subsection{PBiGAMP for joint estimation}
In this section, we explicitly state PBiGAMP \cite{PBiGAMP} for joint estimation in \eqref{bilin} and describe the information contained in the factor nodes of \figref{fig:factor}. To be consistent with the notation used in \cite{PBiGAMP}, we rewrite the random variable dependency corresponding to \eqref{zdepend} in the tensor notation as 
\begin{equation}
\mathsf{z}_m=\sum_{i=1}^{\Nb}\sum_{k=1}^{\Nc}z_{m}^{\left(i,k\right)}\mathsf{b}_i\mathsf{c}_k,
\end{equation} 
where $\Nb= \Np$, $\Nc= \Nrx \Ntx L$, and $z_{m}^{\left(i,k\right)}$ is an element of a third order  tensor given by 
\begin{equation}
z_{m}^{\left(i,k\right)}= \mathbf{\mathbf{G}}_{m,i}\mathbf{\mathbf{A}}_{m,k}.
\label{z_explicit}
\end{equation} 
The output likelihood function in \figref{fig:factor}, denoted by $p_{\mathsf{y}_{m}|\mathsf{z}_{m}}\left(y_{m}|\,z\right)$ is given by
\begin{equation}
p_{\mathsf{y}_{m}|\mathsf{z}_{m}}\left(y_{m}|\,z\right)=\begin{cases}
F\left(\frac{\sqrt{2}\mathrm{sign}\left(\mathrm{Re}\left\{ y_{m}\right\}\right)\mathrm{Re}\left\{ z\right\}}{\sigma}\right)F\left(\frac{\sqrt{2}\mathrm{sign}\left(\mathrm{Im}\left\{ y_{m}\right\}\right)\mathrm{Im}\left\{ z\right\} }{\sigma}\right) & \,q=1\\
\frac{1}{\pi\sigma^{2}}e^{-\frac{\left\Vert y_{m}-z\right\Vert ^{2}}{\sigma^{2}}} & \,q=\infty
\end{cases} ,
\end{equation} 
where $F\left(\cdot \right)$ is the cumulative distribution function of the standard normal distribution.  The sparsity of the vectors $\mathbf{b}$ and $\mathbf{c}$ is incorporated by assuming parametrized Bernoulli-Gaussian distributions for their priors $p_{\boldsymbol{\mathsf{b}}}\left( \mathbf{b} \right) $ and $p_{\boldsymbol{\mathsf{c}}} \left( \mathbf{c} \right)$. For simplicity, it is assumed that each entry of $\boldsymbol{\mathsf{b}}$ is independent of the other and identically distributed as $ p_{\mathsf{b}}$. Similarly, the entries of $\boldsymbol{\mathsf{c}}$ are assumed to be IID, with $ p_{\mathsf{c}}$ as the distribution. Furthermore, the vectors $\boldsymbol{\mathsf{b}}$ and $\boldsymbol{\mathsf{c}}$ are assumed to be independent of each other. Let $\lambda_b$ and $\lambda_c$ denote the sparsity fraction of $\boldsymbol{\mathsf{b}}$ and $\boldsymbol{\mathsf{c}}$. Let $\sigma_b^2$ and $\sigma_c^2$ be the variances of the coefficients corresponding to the non-zero support of the vectors $\boldsymbol{\mathsf{b}}$ and $\boldsymbol{\mathsf{c}}$. With $\delta \left( x \right)$ used to represent the Dirac-delta function, the Bernoulli-Gaussian distributions $p_{\mathsf{b}}$ and $p_{\mathsf{c}}$ can be given as 
\begin{align}
p_{\mathsf{b}}\left(x\right) & =\lambda_{b}\delta\left(x\right)+\left(1-\lambda_{b}\right)\mathcal{N}(0,\sigma_{b}^{2}),\\
p_{\mathsf{c}}\left(x\right) & =\lambda_{c}\delta\left(x\right)+\left(1-\lambda_{c}\right)\mathcal{N}(0,\sigma_{c}^{2}).
\end{align} 
The parameters governing $p_{\mathsf{b}}$ and $p_{\mathsf{c}}$, however, are not known apriori and can be learned by embedding PBiGAMP within the Expectation Maximization (EM) algorithm \cite{EMGAMP}. For a given set of likelihood functions and prior distributions, the vector variance PBiGAMP algorithm \cite{PBiGAMP} to obtain the MMSE estimates of $\mathbf{b}$ and $\mathbf{c}$ in \eqref{bilin} is summarized in \tabref{PbigAmp}. 
\putTable{PbigAmp}{The PBiGAMP Algorithm from \cite{PBiGAMP}
}{\scriptsize
\begin{equation*}
\begin{array}{|lr@{\,}c@{\,}l@{}r|}\hline
  \multicolumn{2}{|l}{\textsf{Definitions:}}&&&\\[-1mm]
  &p_{\mathsf{z}_{m}|\mathsf{p}_{m}\!}\big(z\giv\hat{p};\nu^p\big) 
   &\defn& \frac{p_{\mathsf{y}_m|\mathsf{z}_{m}\!}(y_{m} \giv z) \, \mathcal{N}(z;\hat{p},\nu^p)}{\int_{z'}p_{\mathsf{y}_m|\mathsf{z}_{m}\!}(y_{m} \giv z') \, \mathcal{N}(z';\hat{p},\nu^p)} &\text{(D1)}\\
   
&p_{\mathsf{c}_k|\mathsf{r}_k\!}(\Cr\giv\hat{r};\nu^r) 
	&\defn& \frac{p_{\mathsf{c} \!}(\Cr) \, \mathcal{N}(\Cr;\hat{r},\nu^r)}{\int_{\Cr'}p_{\mathsf{c} \!}(\Cr') \, \mathcal{N}(\Cr';\hat{r},\nu^r)}&\text{(D2)}\\
&p_{\mathsf{b}_i|\mathsf{q}_i\!}(\Br\giv\hat{q};\nu^q) 
	&\defn& \frac{p_{\mathsf{b} \!}(\Br) \, \mathcal{N}(\Br;\hat{q},\nu^q)}{\int_{\Br'}p_{\mathsf{b}\!}(\Br') \, \mathcal{N}(\Br';\hat{q},\nu^q)}&\text{(D3)}\\
	
    \multicolumn{2}{|l}{\textsf{Initializations:}}&&&\\
    &\forall m:
      \hat{s}_{m}(0) &=& 0  & \text{(I1)}\\
  &\forall \iB,\jC: \textsf{choose~} &
  \multicolumn{2}{l}{\hat{\Br}_{\iB}(1), \nu^{\Br}_{\iB}(1), \hat{\Cr}_{\jC}(1), \nu^{\Cr}_{\jC}(1)} &\text{(I2)}\\  
  \multicolumn{2}{|l}{\textsf{for $t=1,\dots T_\textrm{max}$}}&&&\\
   &\forall m,\iB:
   \zmat{m}{\iB}{\nod}{t}{}
   &=& \sum_{\jC=1}^{\Nc} z_{m}\of{\iB,\jC} \hat{c}_{\jC}(t) & \text{(R1)}\\[0.5mm]
   & \forall m,\jC:
   \zmat{m}{\nod}{\jC}{t}{}
   &=& \sum_{\iB=1}^{\Nb} \hat{b}_{\iB}(t) z_{m}\of{\iB,\jC} & \text{(R2)}\\[0.5mm]  
   & \forall m:
   \zmat{m}{\nod}{\nod}{t}{} 
   &=& \sum_{\iB=1}^{\Nb} \hat{b}_{\iB}(t) \zmat{m}{\iB}{\nod}{t}{} 
       \text{~or~} \sum_{\jC=1}^{\Nc} \hat{c}_{\jC}(t) \zmat{m}{\nod}{\jC}{t}{}
       & \text{(R3)}\\[0.5mm]  
   &\forall m:
   \bar{\nu}^p_{m}(t)
   &=& \sum_{\iB=1}^{\Nb} \nu^{\Br}_{\iB}(t) |\zmat{m}{\iB}{\nod}{t}{}|^2   
   + \sum_{\jC=1}^{\Nc} \nu^{\Cr}_{\jC}(t) |\zmat{m}{\nod}{\jC}{t}{}|^2 & \text{(R4)}\\[0.5mm]
   &\forall m:
   \nu^p_{m}(t)
   &=& \bar{\nu}^p_{m}(t)+  \sum_{\iB=1}^{\Nb} \nu^{\Br}_{\iB}(t) \sum_{\jC=1}^{\Nc} \nu^{\Cr}_{\jC}(t)|z_{m}\of{\iB,\jC}|^2 & \text{(R5)}\\[0.5mm] 

   &\forall m: \hat{p}_{m}(t) 
   &=& \zmat{m}{\nod}{\nod}{t}{} - \hat{s}_{m}(t\!-\!1)\bar{\nu}^p_{m}(t)& \text{(R6)}\\[0.5mm] 
   &\forall m: \nu^z_{m}(t) 
   &=& \var\{\mathsf{z}_{m}\giv\mathsf{p}_m\!=\!\hat{p}_{m}(t);\nu^p_{m}(t)\} & \text{(R7)}\\[0.5mm] 
   &\forall m: \hat{z}_{m}(t) 
   &=& \E\{\mathsf{z}_{m}\giv\mathsf{p}_m\!=\!\hat{p}_{m}(t);\nu^p_{m}(t)\} & \text{(R8)}\\[0.5mm] 
   &\forall m: \nu^s_{m}(t) 
   &=& (1 -  \nu^z_{m}(t)/\nu^p_{m}(t) )/\nu^p_{m}(t)  & \text{(R9)}\\[0.5mm] 
   &\forall m: \hat{s}_{m}(t) 
   &=& ( \hat{z}_{m}(t) - \hat{p}_{m}(t))/\nu^p_{m}(t) & \text{(R10)}\\[0.5mm] 
   &\forall \jC: \nu^r_{\jC}(t)
   &=&  \Big( \sum_{m=1}^M \nu^s_{m}(t) |\zmat{m}{\nod}{\jC}{t}{}|^2  \Big)^{-1} & \text{(R11)}\\[1mm] 
   &\forall \jC: \hat{r}_{\jC}(t) 
   &=& \hat{\Cr}_{\jC}(t) + \nu^r_{\jC}(t) \sum_{m=1}^M \hat{s}_{m}(t) \zmat{m}{\nod}{\jC}{t}{*} &\\
   &&&~ -\nu^r_{\jC}(t)\hat{\Cr}_{\jC}(t) \sum_{m=1}^M \nu^s_{m}(t) \sum_{\iB=1}^{\Nb} \nu^{\Br}_{\iB}(t) 
	|\zmatC{m}{\iB}{\jC}{t}{}|^2 & \text{(R12)}\\[0.5mm]
   &\forall \iB: \nu^q_{\iB}(t) 
   &=&  \Big( \sum_{m=1}^M \nu^s_{m}(t) |\zmat{m}{\iB}{\nod}{t}{}|^2  \Big)^{-1} & \text{(R13)}\\[1mm] 
   &\forall \iB: \hat{q}_{\iB}(t) 
   &=& \hat{\Br}_{\iB}(t) + \nu^q_{\iB}(t) \sum_{m=1}^M \hat{s}_{m}(t)   \zmat{m}{\iB}{\nod}{t}{*} &\\
   &&&~ -\nu^q_{\iB}(t)\hat{\Br}_{\iB}(t) \sum_{m=1}^M \nu^s_{m}(t) \sum_{\jC=1}^{\Nc} \nu^{\Cr}_{\jC}(t) 
        |\zmatC{m}{\iB}{\jC}{t}{}|^2  &\text{(R14)}\\[0.5mm]
   &\forall \jC: \nu^{\Cr}_{\jC}(t\!+\!1) 
   &=& \var\{\mathsf{c}_{k}\giv\mathsf{r}_k\!=\!\hat{r}_{\jC}(t); \nu^r_{\jC}(t)\} & \text{(R15)}\\[0.5mm] 
   &\forall \jC: \hat{\Cr}_{\jC}(t\!+\!1) 
   &=& \E\{\mathsf{c}_{k}\giv\mathsf{r}_k\!=\!\hat{r}_{\jC}(t); \nu^r_{\jC}(t)\}& \text{(R16)}\\[0.5mm] 
   &\forall \iB: \nu^{\Br}_{\iB}(t\!+\!1) 
   &=& \var\{\mathsf{b}_{i}\giv\mathsf{q}_i\!=\!\hat{q}_{\iB}(t); \nu^q_{\iB}(t)\}& \text{(R17)}\\[0.5mm] 
   &\forall \iB: \hat{\Br}_{\iB}(t\!+\!1) 
   &=& \E\{\mathsf{b}_{i}\giv\mathsf{q}_i\!=\!\hat{q}_{\iB}(t); \nu^q_{\iB}(t)\}& \text{(R18)}\\[0.5mm] 
   
      \multicolumn{4}{|c}{\textsf{if $\sum_{m=1}^M |\zmat{m}{\nod}{\nod}{t}{} - \zmat{m}{\nod}{\nod}{t\!-\!1}{}|^2 \le \tau_\textrm{stop} \sum_{m=1}^M |\zmat{m}{\nod}{\nod}{t}{}|^2$, \textsf{stop}}}&\text{(R19)}\\
    \multicolumn{2}{|l}{\textsf{end}}&&&\\
    \multicolumn{2}{|l}{\textsf{Output:} \,\, \hat{\mathbf{b}}=\hat{\mathbf{b}} \left(t \right), \hat{\mathbf{c}}=\hat{\mathbf{c}} \left(t \right).}&&&\\\hline
\end{array}
\end{equation*}
}
\par The channel and the phase error vector can be derived using appropriate transformations over the PBiGAMP estimates. The vector $\hat{\mathbf{c}}$ obtained from PBiGAMP is just an estimate of the vectorized version of $\mathbf{C}$, the angle-delay domain representation of the wideband channel in \eqref{block_model_bilin}. Let $\hat{\mathbf{C}} \in \mathbb{C}^{\Nrx \times \Ntx L}$ denote the angle-delay domain estimate of the wideband channel, such that $\mathrm{vec}\left(\hat{\mathbf{C}}\right)=\hat{\mathbf{c}}$. It can be seen from \eqref{z_pre_equation} that $\mathbf{C}$ is just a concatenation of the angle domain representation of all the $L$ taps of the MIMO channel. Therefore, the $\ell^\mathrm{th}$ tap of the antenna domain MIMO channel can be derived as  
\begin{equation}
\hat{\mathbf{H}}\left[\ell\right]=\mathbf{U}_{\Nrx}\left[\hat{\mathbf{C}}_{(\ell\Ntx+1)},\hat{\mathbf{C}}_{(\ell \Ntx+2)},\,.\,.\,.,\hat{\mathbf{C}}_{(\ell\Ntx+\Ntx)}\right]\mathbf{\mathbf{U}^{\ast}}_{\Ntx}.
\label{hest}
\end{equation} 
The vector $\hat{\mathbf{b}}$ derived from PBiGAMP is an estimate of the DFT of the phase error vector $\mathbf{d}\left(\epsilon, \beta \right)$. Estimating the CFO from $\hat{\mathbf{b}}$ is just a single tone frequency estimation problem. As the phase noise is modelled as a Wiener process, we apply the Extended Kalman Filter (EKF) \cite{EKF} over the time domain samples, i.e., $\mathbf{U}^{\ast}_{\Np}\hat{\mathbf{b}}$, to estimate $\epsilon$. 
\par In this work, we exploit the compressibility of the phase error vector in the DFT basis for joint estimation. It is possible, however, to incorporate the statistics of the phase noise by replacing the nodes corresponding to $\mathbf{b}$ in Fig. \ref{fig:factor} with the phase error variables and adding factors corresponding to the phase noise process. In such case, the message passing algorithm must handle the non-linear dependence of $\mathbf{z}$ on the phase errors.     
\subsection{PBiGAMP: From theory to practice} \label{sec:fastimpl}
As seen from \tabref{PbigAmp}, the generic implementation of PBiGAMP is memory and computationally intensive, as it involves repeated operations over a third order tensor in \eqref{z_explicit}. In this section, we provide insights into the key equations of \tabref{PbigAmp} and describe our low complexity implementation for joint estimation. 
\par The equations in PBiGAMP are essentially determined by the belief flows from the factor nodes to the variable nodes, and vice versa. Without loss of generality, we describe the message flow between the variable node $\mathsf{b}_1$ and the factor node $\mathsf{y}_1$ in \figref{fig:factor}. In standard message passing, the message sent from a factor node to $\mathsf{b}_1$ essentially represents the PDF of $\mathsf{b}_1$ presumed by that factor node. In a fully connected factor graph, the node $\mathsf{b}_1$ receives messages from all the $M$ factor nodes ($\left\{ \mathsf{y}_{i}\right\} _{i=1}^{M}$) in addition to its prior distribution $p_{\mathsf{b}}$.
As $\mathsf{b}_1$ receives several beliefs from different factors, the belief sent by $\mathsf{b}_1$ to $\mathsf{y}_1$ is just a normalized product of all the beliefs received by $\mathsf{b}_1$ except the one from $\mathsf{y}_1$. For generic priors and likelihood functions in the factor graph, performing standard message passing can be difficult as the belief flows are flows of PDFs that are functions.
\par PBiGAMP simplifies standard message passing using the central limit theorem (CLT) and the Taylor series approximation. Notice that $\mathsf{y}_1$ receives beliefs from all its neighbouring variable nodes and contains the likelihood function in itself. The belief sent from $\mathsf{y}_1$ to $\mathsf{b}_1$ is computed by multiplying the likelihood function, with all the incoming beliefs to $\mathsf{y}_1$ except the one from $\mathsf{b}_1$, and then integrating over all the random variables except $\mathsf{b}_1$. The multiplication followed by integration essentially yields the PDF of $\mathsf{b}_1$ presumed by the factor node $\mathsf{y}_1$. The integration is often multidimensional and can be difficult to compute. If $\mathsf{y}_1$ depends on a large number of independent variable nodes through a linear function, then the linear combination can be approximated as a Gaussian random variable using the CLT \cite{AMP}. In such case, the variable nodes can send just the mean and variances of the PDFs to the factors and this information is sufficient to compute the mean and variance of the Gaussian random variable, as seen in (R3) and (R5) of \tabref{PbigAmp}. The message sent from $\mathsf{y}_1$ to $\mathsf{b}_1$ can now be computed by multiplying the likelihood at $\mathsf{y}_1$ with the compound Gaussian PDF, and marginalizing the product with respect to $\mathsf{b}_1$. In general, the likelihood function can be non-linear in nature and can yield a complicated PDF of $\mathsf{{b}}_1$. Using a second order Taylor series expansion for the log PDF, PBiGAMP \cite{AMP} simplifies the belief sent from $\mathsf{y}_1$ to $\mathsf{b}_1$ to a Gaussian whose mean and variance can be computed from (R7)-(R9) of \tabref{PbigAmp}. For $q$-bit ADCs, the closed form expressions for the conditional mean and variance in (R7) and (R8) can be found in \cite[Appendix A]{JCD}.  
Unlike standard message passing, PBiGAMP is computationally tractable as the messages contain just the mean and variances of the PDFs.    
\par After several approximate message flows between the factor nodes and the variable nodes, PBiGAMP is expected to converge. The MMSE of $b_1$ is computed as the expectation of the effective marginal, i.e., the normalized product of all the $M$ Gaussian PDFs received by $\mathsf{b}_1$ from the factor nodes and the prior distribution on $\mathsf{b}_1$. Notice that the normalization has to be done to ensure that the PDF integrates to 1. The expectation step for the MMSE of $b_1$ is given in (R18) using (R11) and (R12) as the intermediate steps. Similarly, the expectation for the MMSE of $c_1$ is given in (R16) using (R13) and (R14) as the intermediate steps. Thus, PBiGAMP provides estimates of the vectorized sparse channel and the DFT of the phase error vector.   
\par The generic implementation of PBiGAMP is computationally expensive primarily due to (R1), (R2), (R5), (R12) and (R14) in \tabref{PbigAmp}. It can be verified that each of  these operations have a complexity of $O \left( \Nrx ^2 \Np^2 \Ntx L \right)$ for a single PBiGAMP iteration to perform joint estimation. For instance, (R1) requires computing the scalar $\zmat{m}{\iB}{\nod}{t}{}$ for every $m\in \mathcal{I}_{M}$ and $i \in \mathcal{I}_{\Nb}$, Therefore, (R1) demands $O(M\Nb \Nc)$ computations, as each scalar computation requires $\Nc$ multiplications. As $M=\Nrx \Np$, $\Nb=\Np$ and $\Nc= \Nrx\Ntx L$ for the joint estimation, (R1) has a complexity of $O \left( \Nrx ^2 \Np^2 \Ntx L \right)$ for every PBiGAMP iteration. By exploiting the structure in the joint estimation problem, the complexity of PBiGAMP can be significantly reduced.
\par We describe our fast implementation of PBiGAMP for the joint estimation in the following sub-sections. The $t^{\mathrm{th}}$ iteration variables of PBiGAMP are defined as $\hat{\mathbf{b}}\left(t\right) \in \mathbb{C}^{\Np \times 1}$, $\hat{\mathbf{C}}\left(t\right) \in \mathbb{C}^{\Nrx \times \Ntx L}$ and $\hat{\mathbf{Z}}\left(t\right) \in \mathbb{C}^{\Nrx \times \Np}$, such that $\hat{b}_{i}\left(t\right),\hat{c}_{k}\left(t\right)$ and $\hat{z}_{m}\left(t\right)$ are the $i^{\mathrm{th}},k^{\mathrm{th}}$ and $m^{\mathrm{th}}$ entries of $\hat{\mathbf{b}}\left(t\right), \mathrm{vec} ( \hat{\mathbf{C}}(t) )$ and $\mathrm{vec} ( \hat{\mathbf{Z}}(t))$. 
\subsubsection{Operations $\left(\mathrm{R}1\right)$-$\left(\mathrm{R}3\right)$}    
For an $i_o \in \mathcal{I}_{\Np}$, we have 
\begin{align}
\hat{z}_{m}^{\left( i_o,\ast \right)}{(t)} &= \sum_{k=1}^{\Nc}z_{m}^{\left(i_{o},k\right)}\hat{c}_{k}(t)\\
&= \sum_{i=1}^{\Nb}\sum_{k=1}^{\Nc}z_{m}^{\left(i,k\right)}e_{i}^{i_{o},\Np}\hat{c}_{k}(t) \\
&= \mathrm{vec}_m \left( \mathbf{U}_{\Nrx} \hat{\mathbf{C}}(t) \mathbf{F} \mathrm{diag} ( \mathbf{U}^{\ast}_{{\Np}} \mathbf{e}^{i_{o},\Np}) \right)\\
& = \mathrm{vec}_m \left( \mathbf{U}_{\Nrx} \hat{\mathbf{C}}(t) \mathbf{F} \mathrm{diag} ( \mathbf{U}^{\ast}_{{\Np}_{(i_o)}} ) \right),
\label{r1_simp}
\end{align}
where the compact form in \eqref{r1_simp} is obtained by going back from the tensor formulation to the original bilinear model in \eqref{z_pre_equation}. Likewise, for a $k_o \in \mathcal{I}_{\Nc}$, we have 
\begin{align}
\hat{z}_{m}^{\left( \ast, k_o \right)}{(t)} &= \sum_{i=1}^{\Nb}\sum_{k=1}^{\Nc}z_{m}^{\left(i,k\right)}\hat{b}_{i}(t)e_{k}^{k_{o},\Nc} \\
& = \mathrm{vec}_m \left( \mathbf{U}_{\Nrx} \mathbf{e}^{r_o,\Nrx} (\mathbf{e}^{c_o,\Ntx L})^T  \mathbf{F} \mathrm{diag} ( \mathbf{U}^{\ast}_{\Np} \hat{\mathbf{b}}(t) ) \right)\\
&=  \mathrm{vec}_m \left( {\mathbf{U}_{\Nrx}}_{(r_o)} \mathbf{F}^{(c_o)} \mathrm{diag} ( \mathbf{U}^{\ast}_{\Np} \hat{\mathbf{b}}(t) ) \right),
\label{r2_simp}
\end{align}
where $(r_o,c_o)$ correspond to the row and column of the $\Nrx \times \Ntx L$ matrix version of $\mathbf{e}^{k_{o},\Nc}$, with 
\begin{align}
k_o & = (c_o-1)\Nrx +r_o.
\label{rowcol}
\end{align} 
Similarly, (R3) can be computed as
\begin{equation}
\hat{z}_{m}^{\left( \ast, \ast \right)}{(t)}= \mathrm{vec}_m \left( \mathbf{U}_{\Nrx} \hat{\mathbf{C}}(t) \mathbf{F} \mathrm{diag} ( \mathbf{U}^{\ast}_{\Np} \hat{\mathbf{b}}(t) ) \right).
\label{zmhatstst}
\end{equation}
To evaluate \eqref{zmhatstst}, the product $\hat{\mathbf{C}}(t) \mathbf{F}$ can be found using $\Nrx \Ntx L \Np$ computations. Using the FFT over the resultant product, 
$\mathbf{U}_{\Nrx}\hat{\mathbf{C}}(t) \mathbf{F}$ can be computed with an additional complexity of $O(\Np \Nrx \mathrm{log} \Nrx)$. Finally, the complexity to multiply the resultant $\Nrx \times \Np$ matrix with the IFFT of $\hat{\mathbf{b}}(t)$ is $O(\Np \mathrm{log} \Np)+\Nrx \Np$. Therefore, the computational complexity of \eqref{zmhatstst} is $O(\Nrx\Ntx\Np L)$, unlike $O(\Nrx^2\Np^2\Ntx L)$ of the generic implementation using (R1)-(R3).
\subsubsection{Operations $\left(\mathrm{R}4\right),\left(\mathrm{R}5\right)$}
It can be noticed from \eqref{r1_simp} that $\left|\hat{z}_{m}^{(i,\ast)}(t)\right|$ is invariant with respect to $i$ and is given by   
\begin{equation}
\left|\hat{z}_{m}^{(i,\ast)}(t)\right|=\frac{1}{\sqrt{\Np}}\mathrm{vec}_{m}\left(\left|\mathbf{U}_{\Nrx}\hat{\mathbf{C}}(t)\mathbf{F}\right|\right).
\label{mag_zim}
\end{equation}
Using the invariance property in \eqref{mag_zim}, a compact version of the first summand in $\left(\mathrm{R}4\right)$ can be expressed as  
\begin{equation}
\sum_{i=1}^{\Np} v_{i}^{b}\left|\hat{z}_{m}^{(i,\ast)}(t)\right|^{2}=\frac{\sum_{i=1}^{\Np}v_{i}^{b}}{\Np}\mathrm{vec}_{m}\left(\left|\mathbf{U}_{\Nrx}\hat{\mathbf{C}}(t)\mathbf{F}\right|^{2}\right).
\end{equation}
 For the second summand in (R4), it can be shown from \eqref{r2_simp} that 
\begin{align}
\left|\hat{z}_{m}^{(\ast,k_{o})}(t)\right|&=\frac{1}{\sqrt{\Nrx}}\mathrm{vec}_{m}\left(\left|\mathbf{F}^{(c_{o})}\mathrm{diag}\left({\mathbf{U}_{\Np}^{\ast}}\hat{\mathbf{b}}(t)\right)\right|\otimes\mathbf{a}_{\Nrx}(0)\right).
\label{mag_zjm}
\end{align}
Furthermore, as $c_o$ denotes the column number corresponding to $k_o$ (see \eqref{rowcol}), $\left|\hat{z}_{m}^{(\ast,k)}\left(t\right)\right|$ is invariant $\forall k\,\in \mathcal{I}_{a\Nrx}\setminus\mathcal{I}_{(a-1)\Nrx}$, where $a \in \mathcal{I}_{\Ntx L}$. To use this invariance for efficient computation of the second summand in (R4), we define a row vector $\boldsymbol{\mu}^c(t) \in \mathbb{R}^{1 \times \Ntx L}$ containing the column-wise mean corresponding to $\Nrx \times \Ntx L$ matrix version of $\left\{ v_{k}^{c}(t)\right\} _{k\in \mathcal{I}_{\Nc}}$ as 
\begin{equation}
{\mu}^{c}_{n}(t)=\frac{1}{\Nrx}\underset{k\in\mathcal{I}_{n\Nrx}\setminus \mathcal{I}_{(n-1)\Nrx}}{\sum v_{k}^{c}(t)}.
\end{equation} 
With some algebraic manipulation, the second summand in (R4) can be simplified as
\begin{equation}
\sum_{k=1}^{\Nc}v_{k}^{c}(t)\left|\hat{z}_{m}^{(\ast,k)}(t)\right|^{2}= \mathrm{vec}_m \left[ \left(\boldsymbol{\mu}^{c}(t)\left|\mathbf{F}\mathrm{diag}(\mathbf{U}_{\Np}^{\ast}\hat{\mathbf{b}}(t))\right|^{2}\right)\otimes\mathbf{a}_{\Nrx}(0) \right].
\end{equation}
To simplify the computations involved in (R5), we expand the summand using \eqref{z_explicit} as
\begin{align}
\sum_{i=1}^{\Nb} \sum_{k=1}^{\Nc}v_{i}^{b}(t)v_{k}^{c}(t)\left|z_{m}^{(i,k)}\right|^{2} &= \sum_{i=1}^{\Nb} \sum_{k=1}^{\Nc}v_{i}^{b}(t)v_{k}^{c}(t)\left|\mathbf{G}_{m,i}\mathbf{A}_{m,k}\right|^{2}
\label{r5_pre1}\\
&= \frac{\sum_{i=1}^{\Np}v_{i}^{b}(t)}{\Np}\sum_{k=1}^{\Nc}v_{k}^{c}(t)\left|\mathbf{A}_{k,m}^{T}\right|^{2} ,
\label{r5_pre2}
\end{align}
where \eqref{r5_pre2} follows from \eqref{r5_pre1} as $\left| \mathbf{G}_{m,i} \right| =\frac{1}{\sqrt{\Np}}, \, \forall m,i$. Besides, as $\mathbf{A}^T = \mathbf{F} \otimes \mathbf{U}_{\Nrx}$, the entries of $\left| \mathbf{A}^T \right|$ are invariant within blocks of size $\Nrx \times \Nrx$ . With arguments similar to the simplifications involved in the second summand of (R4), (R5) can be efficiently evaluated as
\begin{equation}
\sum_{i=1}^{\Nb} \sum_{k=1}^{\Nc}v_{i}^{b}(t)v_{k}^{c}(t)\left|z_{m}^{(i,k)}\right|^{2} = \frac{\sum_{i=1}^{\Np}v_{i}^{b}(t)}{\Np} \mathrm{vec}_m\left[\left(\boldsymbol{\mu}^{c}(t)\left|\mathbf{F}\right|^{2}\right)\otimes\mathbf{a}_{\Nrx}(0)\right]. 
\label{r5_final}
\end{equation} 
\subsubsection{Operations $\left(\mathrm{R}11\right),\left(\mathrm{R}13\right)$}
From \eqref{mag_zjm}, it can be observed that $\left|\hat{z}_{m}^{(\ast,k)}(t)\right|^2$ is fixed for 
$m \in  \mathcal{I}_{a\Nrx}\setminus \mathcal{I}_{(a-1)\Nrx}$ and $k \in \mathcal{I}_{b\Nrx}\setminus \mathcal{I}_{(b-1)\Nrx} $, where $a \in \mathcal{I}_{\Np}$ and $b \in \mathcal{I}_{ \Ntx L }$. The invariance of $\left|\hat{z}_{m}^{(\ast,k)}(t)\right|^2$ with respect to $m$ and $k$ arise because of the kronecker product with $\mathbf{a}_{\Nrx} \left(0\right)$  and the column invariance in \eqref{rowcol}. To exploit this property in computing $v^r_k (t)$ of (R11), we construct a vector $\boldsymbol{\mu}^z(t) \in \mathbb{R}^{\Np \times 1}$ to contain the column-wise mean corresponding to the $\Nrx \times \Np$ matrix version of $\left\{ v_{m}^{s}(t)\right\} _{m\in\mathcal{I}_{\Np \Nrx }}$ , i.e.,
\begin{equation}
{\mu}_{k}^{z}(t)=\frac{1}{\Nrx}\underset{m\in \mathcal{I}_{k\Nrx} \setminus \mathcal{I}_{(k-1)\Nrx}}{\sum v_{m}^{s}(t)}.
\label{mu_z}
\end{equation}
With the above definitions, a simplified version of $\left(v_{j}^{r}(t)\right)^{-1}$ in (R11) can be given as 
\begin{equation}
\sum_{m=1}^{M}v_{m}^{s}(t)\left|\hat{z}_{m}^{(\ast,k)}\right|^{2} = \mathrm{vec}_{k}\left[\left(\left|\mathbf{F}\mathrm{diag}(\mathbf{U}_{\Np}^{\ast}\hat{\mathbf{b}})\right|^{2}\boldsymbol{\mu}^{z}(t)\right)\otimes\mathbf{a}_{\Nrx}(0)\right]. 
\label{r11_simp}
\end{equation}
From \eqref{mag_zim}, we rewrite $\left(v_{i}^{q}(t)\right)^{-1}$ in (R13) as
\begin{align*}
\left(v_{i}^{q}(t)\right)^{-1} & = \sum_{m=1}^{M}v_{m}^{s}\mathrm{vec}_{m}\left(\left|\mathbf{U}_{\Nrx}\mathbf{\hat{C}}(t)\mathbf{F}\right|^{2}\right)\\
&=\mathbf{v}^{s}\mathrm{vec}\left(\left|\mathbf{U}_{\Nrx}\mathbf{\hat{C}}(t)\mathbf{F}\right|^{2}\right), \hfill \,\,\,\,\,\,\,\,\, \forall i \in \mathcal{I}_{\Np}.
\end{align*}
\subsubsection{Operations $\left(\mathrm{R}12\right),\left(\mathrm{R}14\right)$}
We define $\hat{\mathbf{F}}(t)=\mathbf{F} \mathrm{diag}\left(\mathbf{U}^{\ast}_{\Np} \hat{\mathbf{b}}(t)\right)$ and consider the term in the second summand of (R12) for $k=k_o$. From \eqref{r2_simp} and \eqref{rowcol}, we have
\begin{equation}
\sum_{m=1}^{M}\hat{s}_{m}(t)\hat{z}_{m}^{(\ast,k_{o})}(t)^{\ast}=\sum_{m=1}^{M}\hat{s}_{m}(t)\mathrm{vec}_{m}\left[\overline{\mathbf{U}}_{\Nrx(r_{o})}\overline{\hat{\mathbf{F}}(t)}^{(c_{o})}\right].
\label{r_12pre1}
\end{equation} 
With $\hat{\mathbf{S}}(t) \in \mathbb{C}^{\Nrx \times \Np}$ defined such that $\hat{s}_{m}(t)=\mathrm{vec}_m\left( \hat{\mathbf{S}}(t) \right)$, \eqref{r_12pre1} can be expressed as,
\begin{align}
\sum_{m=1}^{M}\hat{s}_{m}(t)\hat{z}_{m}^{(\ast,k_{o})}(t)^{\ast}&= \left\langle \hat{\mathbf{S}}(t), \mathbf{U}_{\Nrx(r_{o})}\hat{\mathbf{F}}(t)^{(c_{o})} \right\rangle \\
&=\left(\mathbf{U}_{\Nrx(r_{o})}\right)^{\ast}\mathbf{\hat{S}}(t)\left(\hat{\mathbf{F}}(t)^{(c_{o})}\right)^{\ast}\\
&= \mathrm{vec}_{k_o}\left(\mathbf{U}_{\Nrx}^{\ast}\mathbf{\hat{S}}(t)\hat{\mathbf{F}(t)}^{\ast}\right).
\end{align}
The term in the third summand of (R12) can be given by
\begin{align*}
\sum_{m=1}^{M}v_{m}^{s}(t)\sum_{i=1}^{\Np}v_{i}^{b}(t)\left|z_{m}^{(i,k)}\right|^{2} &=\sum_{m=1}^{M}v_{m}^{s}(t)\frac{\sum_{i=1}^{\Np}v_{i}^{b}(t)\left|\mathbf{A}_{m,k}\right|^{2}}{\Np}\\
&=\left(\frac{\sum_{i=1}^{\Np}v_{i}^{b}(t)}{\Np}\right)\mathrm{vec}_k \left( \left|\mathbf{A}^{T}\right|^{2}\mathbf{v}^{s} \right).
\end{align*}
Once again exploiting invariance within $\left|\mathbf{A}^{T}\right|$, we efficiently compute the third summand in (R12) as 
\begin{equation}
\sum_{m=1}^{M}v_{m}^{s}(t)\sum_{i=1}^{\Np}v_{i}^{b}(t)\left|z_{m}^{(i,k)}\right|^{2}=\left(\frac{\sum_{i=1}^{\Np}v_{i}^{b}(t)}{\Np}\right)\mathrm{vec}_k \left[\left(\left|\mathbf{F}\right|^{2}\boldsymbol{\mu}^{z}(t)\right)\otimes\mathbf{a}_{\Nrx}(0)\right].
\end{equation}
The term in the second summand of (R14) can be rewritten using \eqref{r1_simp} as
\begin{equation}
\sum_{m=1}^{M} \hat{s}_{m}(t) \hat{z}_{m}^{{(i,\ast)}{\ast}}=\left\langle \hat{\mathbf{S}}(t),\mathbf{U}_{\Nrx}\hat{\mathbf{C}}(t)\mathbf{F}\mathrm{diag}(\mathbf{U}^{\ast}_{{\Np}_{(i)}})\right\rangle. 
\label{r14_t1}
\end{equation}
For fast implementation of \eqref{r14_t1}, we first compute the column wise inner product between $\hat{\mathbf{S}}(t)$ and $\mathbf{U}_{\Nrx}\hat{\mathbf{C}}(t)\mathbf{F}$ and then perform a Fast Fourier Transform (FFT). We construct $\mathbf{g} \in \mathbb{C}^{\Np \times 1}$, such that $\mathbf{g}_k= \left\langle \left(\hat{\mathbf{S}}(t)\right)_{(k)},\left(\mathbf{U}_{\Nrx}\hat{\mathbf{C}}(t)\mathbf{F}\right)_{(k)}\right\rangle $. Hence, the second summand in (R14) simplifies to 
\begin{equation}
\sum_{m=1}^{M} \hat{s}_{m}(t) \hat{z}_{m}^{{(i,\ast)}{\ast}}= \mathrm{vec}_i \left( \mathbf{U}_{{\Np}} \mathbf{g} \right).
\label{r14_sum2}
\end{equation}
The term in the third summand of (R14) can be expressed as
\begin{align}
\sum_{m=1}^{M}\sum_{k=1}^{\Nc}v_{m}^{s}(t)v_{k}^{c}(t)\left|z_{m}^{(i,k)}\right|^{2}&=\sum_{m=1}^{M}\sum_{k=1}^{\Nc}v_{m}^{s}(t)v_{k}^{c}(t)\left|\mathbf{G}_{m,i}\mathbf{A}_{m,k}\right|^{2}\\
&=\frac{1}{\Np}\sum_{m=1}^{M}v_{m}^{s}(t)\sum_{k=1}^{\Nc}v_{k}^{c}(t)\left|\mathbf{A}_{k,m}^{T}\right|^{2}.
\label{r14_sum3_pre1}
\end{align}
Using the compact form of $\sum_{k=1}^{\Nc}v_{k}^{c}(t)\left|\mathbf{A}_{k,m}^{T}\right|^{2}$ from \eqref{r5_final}, we rewrite \eqref{r14_sum3_pre1} as
\begin{align}
\sum_{m=1}^{M}\sum_{k=1}^{\Nc}v_{m}^{s}(t)v_{k}^{c}(t)\left|z_{m}^{(i,k)}\right|^{2} & =\frac{1}{\Np}\sum_{m=1}^{M}v_{m}^{s}(t)\mathrm{vec}_{m}\left[\left(\boldsymbol{\mu}^{c}(t)\left|\mathbf{F}\right|^{2}\right)\otimes\mathbf{a}_{\Nrx}(0)\right].
\end{align}
Furthermore, exploiting the kroenecker product with $\mathbf{a}_{\Nrx}(0)$, we have 
\begin{equation}
\sum_{m=1}^{M}\sum_{k=1}^{\Nc}v_{m}^{s}(t)v_{k}^{c}(t)\left|z_{m}^{(i,k)}\right|^{2}  =\frac{\Nrx}{\Np}\boldsymbol{\mu}^{c}(t)\left|\mathbf{F}\right|^{2}\boldsymbol{\mu}^{z}(t).
\end{equation}
\subsubsection{Complexity of the simplified operations}
Using the complexity of matrix multiplications and FFTs, the complexity of our simplifications is  summarized in \tabref{compCost}. 
It can be noticed that the overall complexity of PBiGAMP using our implementation is $O \left( \Nrx \Np \Ntx L \right)$, thereby achieving a speedup factor of $\Np\Nrx$ compared to the generic implementation.  
After all possible simplifications of PBiGAMP for joint estimation, training design is  possibly the only frontier that can be exploited to further reduce the computational complexity.  
\putTable{compCost}{Complexity of a single PBiGAMP iteration using a fast implementation.}
{
 \color{\longcolor}
 \renewcommand{\arraystretch}{1}
 \begin{tabular}{|c|c|}
\hline 
Operation & Complexity\tabularnewline
\hline 
\hline 
$\mathbf{U}_{\Nrx}\hat{\mathbf{C}}\left(t\right)\mathbf{F}$ & $O\left(\Nrx\Ntx L\Np\right)$\tabularnewline
\hline 
$\mathbf{U}_{\Np}^{\ast}\hat{\mathbf{b}}\left(t\right)$ & $O\left(\Np\mathrm{log} \Np \right)$\tabularnewline
\hline 
(R3), (R13) & $O\left(\Nrx\Np\right)$\tabularnewline
\hline 
(R4), (R5), (R11) & $O\left(\Ntx L\Np\right)$\tabularnewline
\hline 
(R12), (R14) & $O\left(\Nrx\Ntx L\Np\right)$ \tabularnewline
\hline 
\end{tabular}
\bigskip
}
\subsection{Insights into training design} \label{sec:circulantTrn}
In this section, we explain how a reasonable training solution that allows a low complexity implementation does not permit joint estimation. Structured training blocks that aid fast transforms can reduce the complexity of PBiGAMP operations involving multiplications with the training matrix. Furthermore, such blocks also occupy a lower memory footprint relative to unstructured ones of the same dimension. For example, training blocks that contain circulantly shifted rows of a fixed Zadoff-Chu (ZC) sequence were proposed in \cite{mo2016channel} for fast channel estimation using EM-GAMP. Similar to EM-GAMP algorithm, circulant training matrices also aid fast matrix multiplications in PBiGAMP. For instance, the computation of $\hat{\mathbf{C}}(t) \mathbf{F}$ in \tabref{compCost} can be accelerated in every PBiGAMP iteration for the joint estimation. Shifted ZC training blocks, however, result in a continuum of optimal solutions for bilinear optimization problem. We illustrate the ``CFO propagation effect'' to show the trade-off between fast message passing and identifiability in the joint estimation problem. 
\par To explain the CFO propagation effect, we consider $\Np = \Ntx L$ pilots and a circulant training matrix $\mathbf{T}$ as per \cite{mo2016channel}, for a narrowband system, i.e., $L=1$. Because ${\mathbf{T}}^{\ast}$ is also circulant, the eigenvectors of ${\mathbf{T}}^{\ast}$ are the columns of the DFT matrix $\mathbf{U}_{\Ntx}$. Let $\mathbf{\Lambda}_{\mathbf{T}}$ be a diagonal matrix containing the conjugated eigenvalues of $\mathbf{T}^{\ast}$, such that ${\mathbf{T}}^{\ast} \mathbf{U}_{\Ntx}=\mathbf{U}_{\Ntx}\mathbf{\Lambda}_{\mathbf{T}}^{\ast}$. With $\mathbf{\Lambda}_{\epsilon}$ defined as $\mathrm{diag}\left( \mathbf{d}\left( \epsilon, \beta \right)\right)$, the noiseless unquantized received block in \eqref{z_pre_equation} can be given as 
\begin{align}
\mathbf{Z} & = \mathbf{U}_{\Nrx}\mathbf{C}\mathbf{U^{\ast}}_{\Ntx}\mathbf{T}\mathbf{\Lambda}_{\epsilon}\\
& = \mathbf{U}_{\Nrx}\mathbf{C} \mathbf{\Lambda}_{\mathbf{T}} \mathbf{U^{\ast}}_{\Ntx}\mathbf{\Lambda}_{\epsilon}.
\label{cfoprop_z}
\end{align} 
For our analysis, we assume that there is no phase noise, i.e., $\beta=0$ and choose the CFO ($\epsilon$) to be an integer multiple of ${2\pi}/{\Np}$. We define the integer $d={\Np \epsilon}/{2 \pi}$ and interpret $\mathbf{\Lambda}_{\epsilon}$ as a matrix containing the eigenvalues of the $d$ circulant delay matrix $\mathbf{J}_{d} \in \mathbb{C}^{\Np \times \Np}$, i.e., $\mathbf{J}_{d}\mathbf{U}_{\Ntx}=\mathbf{U}_{\Ntx}\mathbf{\Lambda}_{\epsilon}$. As $\mathbf{\Lambda}^{\ast}_{\epsilon}=\mathbf{\Lambda}_{-\epsilon}$, it can be shown that $\mathbf{J}_{\Ntx-d}\mathbf{U}_{\Ntx}=\mathbf{U}_{\Ntx}\mathbf{\Lambda}^{\ast}_{\epsilon}$. Furthermore, as $\mathbf{J}_{\Ntx-d}$ is a real matrix and $\overline{\mathbf{U}}_{\Ntx}=\mathbf{U}^{\ast}_{\Ntx}$, we have $\mathbf{U}^{\ast}_{\Ntx}\mathbf{\Lambda}_{\epsilon}=\mathbf{J}_{\Ntx-d}\mathbf{U}^{\ast}_{\Ntx}$. It follows from \eqref{cfoprop_z} that  
\begin{equation}
\mathbf{Z}=\mathbf{U}_{\Nrx}\mathbf{C} \mathbf{\Lambda}_{\mathbf{T}} \mathbf{J}_{\Ntx-d} \mathbf{U^{\ast}}_{\Ntx}.
\label{cfoprop_z2}
\end{equation}
Comparing \eqref{cfoprop_z2} with \eqref{cfoprop_z}, it can be observed that the beamspace matrix $\mathbf{C}$ and a CFO of $2 \pi d / \Np$, result in the same received samples as the beamspace matrix $\mathbf{C} \mathbf{\Lambda}_{\mathbf{T}} \mathbf{J}_{\Ntx-d}$ and zero CFO. We call this as the CFO propagation effect and study its impact on the identifiability of the channel and the CFO in joint estimation.
\par Now, we show that training blocks consisting of circulantly shifted ZC sequences propagate the CFO into the channel in an inseparable manner. From the perfect autocorrelation property of ZC sequences, it can be concluded that the diagonal entries of $\mathbf{\Lambda}_{\mathbf{T}}$ have constant modulus. In such case, $\mathbf{\Lambda}^{-1}_{\mathbf{T}}$ is well defined and \eqref{cfoprop_z2} can be rewritten as 
\begin{align}
\mathbf{Z} & = \mathbf{U}_{\Nrx}\mathbf{C}\mathbf{\Lambda}_{\mathbf{T}}\mathbf{J}_{\Ntx-d}\mathbf{\Lambda}_{\mathbf{T}}^{-1}\mathbf{\Lambda}_{\mathbf{T}}\mathbf{U^{\ast}}_{\Ntx}\\
 & = \mathbf{U}_{\Nrx}\underbrace{\mathbf{C}\mathbf{\Lambda}_{\mathbf{T}}\mathbf{J}_{\Ntx-d}\mathbf{\Lambda}_{\mathbf{T}}^{-1}}_{\overset{\Delta}{=}\mathbf{C}(\epsilon)}\mathbf{U^{\ast}}_{\Ntx} \mathbf{T}\\
 & = \mathbf{U}_{\Nrx} \mathbf{C}(\epsilon) \mathbf{U^{\ast}}_{\Ntx} \mathbf{T} \mathbf{\Lambda}_0. 
\label{cfoprop_z3}
\end{align}
Notice that $\mathbf{\Lambda}_0$ in \eqref{cfoprop_z3} is just an identity matrix, also interpreted as a zero CFO perturbation. We compare \eqref{cfoprop_z3} with \eqref{cfoprop_z} and define a beamspace matrix matrix $\mathbf{C}(\epsilon)=\mathbf{C}\mathbf{\Lambda}_{\mathbf{T}}\mathbf{J}_{\Ntx-d}\mathbf{\Lambda}_{\mathbf{T}}^{-1}$. It can be noticed that $\mathbf{C}(\epsilon)$ has the same sparsity as that of $\mathbf{C}$, as permutation and scaling operations determined by $\mathbf{J}_{\Ntx-d}$ and $\mathbf{\Lambda}_{\mathbf{T}}$ preserve sparsity. Therefore, if $\left(\mathbf{C}, \epsilon \right)$ is a solution to the joint estimation problem, $\left(\mathbf{C}(\epsilon), 0 \right)$ is also a solution. In fact, any frequency that is an integer multiple of $2 \pi /\Np $ can be propagated into the channel matrix so that $\left(\mathbf{C}(2 \pi k / \Np ), \epsilon - 2 \pi k / \Np\right)$ is a solution for every $k \in \mathcal{I}_{\Np}$. Therefore, the circulantly shifted ZC training that aids fast message passing results in a continuum of optimal solutions for the joint estimation problem.   
\section{Simulations}
In this section, we provide simulation results for the proposed bilinear message passing based joint CFO and channel estimation algorithm. We consider a hardware architecture in \figref{fig:architect} that uses a uniform linear array of antennas and choose $\Ntx=\Nrx=32$. In this work, we assume that the receiver is equipped with one-bit ADCs, i.e., $q = 1$. To obtain a performance benchmark for one-bit receivers, we evaluate our algorithm for full resolution receivers, i.e., $q=\infty$, although they may not be practical at large bandwidths due to high power consumption. We consider a mmWave carrier frequency of $38 \, \mathrm{GHz}$ and an operating bandwidth of $W=100 \, \mathrm{MHz}$ \cite{bandwidthref}, which corresponds to a symbol duration of $T=10\, \mathrm{ns}$.
\par We describe the simulation parameters of the clustered mmWave channel model in \eqref{wbchannel}. We consider $L=16$ taps to model the mmWave channel in the $100\, \mathrm{MHz}$ bandwidth. Such model is good enough for RMS delay spreads upto $33 \, \mathrm{ns}$, assuming $L=5W\tau_{\mathrm{rms}}$. Our assumption is reasonable as $95\%$ of the measured RMS delay spreads at $38 \, \mathrm{GHz}$ were found to be less than $37.8\, \mathrm{ns}$ \cite{channelmeas}. The measurements in \cite{channelmeas} were made in a UMi-LoS environment using a resolution of $2\, \mathrm{ns}$. We assume $N_{\mathrm{cs}}=4$ clusters, each comprising of $10$ rays and the complex ray gains as IID standard normal random variables. Furthermore, the AoAs and AoDs of the rays within a cluster are chosen from a laplacian distribution corresponding to an angle spread of $15^{\circ}$. 
The wideband channel is scaled so that $\mathbb{E}\left[\sum_{\ell}\left\Vert \mathbf{H}\left[\ell\right]\right\Vert _{\text{F}}^{2}\right]=\mathbb{E}\left[ \left\Vert \mathbf{C} \right\Vert _{\text{F}}^{2} \right]=\Nrx\Ntx$, where the expectation is taken across several channel realizations. The channel matrix generated with the aforementioned  parameters is practical at mmWave and the angle-delay domain representation, i.e., $\mathbf{C}$, can be verified to be approximately sparse. Our joint estimation algorithm does not require any knowledge about the sparsity order of $\mathbf{C}$ and learns it using the EM algorithm. 
\par Our PBiGAMP based approach exploits compressibility of the phase error vector in the Fourier basis. To evaluate the worst case performance of our algorithm, we choose a CFO that is maximally off grid and within the practical limits. As the resolution of the DFT grid is $2 \pi / \Np$, we choose $\epsilon= 45 \pi / 1024$ for a training block of $\Np=1024$ pilots. The corresponding analog domain CFO  can be verified to be $2.2 \, \mathrm{MHz}$, which is about $58 \, \mathrm{ppm}$ of the carrier frequency. We set $\beta_{\mathrm{tx}}=\beta_{\mathrm{rx}}= 0.047\, \mathrm{rad}$ so that the standard deviation of the effective Wiener phase noise process is $\beta = 0.067 \, \mathrm{rad}$. These parameters translate to a phase noise level of about $-85\, \mathrm{dBc}/ \mathrm{Hz}$ at $1\, \mathrm{MHz}$ offset for each of the TX and RX oscillators \cite{lorentzian}, and meet the specifications of a $38 \, \mathrm{GHz}$ oscillator \cite{VCOspecs}. 
\par The performance of our joint estimation algorithm is evaluated using the Normalised Mean Square Error (NMSE) of the channel estimate and the mean square error (MSE) of the CFO estimate $\epsilon$. For a given SNR, the variance of the IID Gaussian noise, i.e., $\sigma^2$ in \eqref{basicmodel}, is chosen such that 
\begin{equation}
\mathrm{SNR}=10\mathrm{log}_{10} \left( \frac{\left\Vert \mathbf{T}\right\Vert _{\mathrm{F}}^{2}}{\Np\sigma^{2}} \right).
\end{equation}
For our simulations, we consider training blocks comprising of IID QPSK entries, IID Gaussian entries, and shifted ZC sequences proposed in \cite{mo2016channel}. For mmWave systems, the IID QPSK training is more practical than the IID Gaussian one, as it can be generated using a TX architecture that is as simple as  analog-beamforming with 2-bit phase shifters. 
\subsection{NMSE of the channel estimate}
Due to the bilinear nature of the problem, we can only estimate the wideband channel or equivalently $\mathbf{C}$ upto a scale factor. Furthermore, any positive amplification of $\mathbf{C}$ results in the same received block in one-bit receivers at high SNR. Therefore, the NMSE of the channel estimate is defined as 
\begin{equation}
\mathrm{NMSE}=\mathbb{E}\left[\frac{{\left\Vert \mathbf{C}- \gamma \hat{\mathbf{C}}\right\Vert^2 _{\mathrm{F}}}}{{\left\Vert \mathbf{C}\right\Vert^2 _{\mathrm{F}}}} \right],
\end{equation}
where $\gamma$ is a scalar such that $\gamma=\underset{a}{\mathrm{arg\,min}}\left\Vert \mathbf{C}-a\hat{\mathbf{C}}\right\Vert _{\mathrm{F}}$ for a given $\mathbf{C}$ and $\hat{\mathbf{C}}$. The matrix $\gamma \hat{\mathbf{C}}$ can be considered as the normalized angle-delay domain estimate of the wideband channel. The function $\mathbb{E}[.]$ denotes the empirical expectation and is taken across several channel and training realizations. For a given realization of the channel and its estimate, we define the Normalized Squared Error (NSE) as $\left\Vert \mathbf{C}- \gamma \hat{\mathbf{C}}\right\Vert^2 _{\mathrm{F}}/{\left\Vert \mathbf{C}\right\Vert^2 _{\mathrm{F}}}$. 
\par
\begin{figure}[h]
\centering
\includegraphics[trim=1cm 6cm 2cm 7.5cm,clip=true,width=0.55 \textwidth]{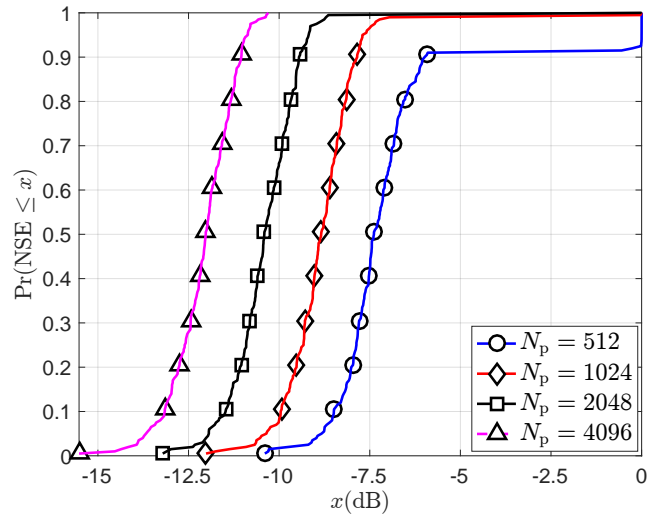}
  \caption{The empirical CDF of the Normalized Squared Error of the channel estimate obtained using an IID QPSK training, at a SNR of $0 \, \mathrm{dB}$ in a one-bit receiver. The reconstruction performance in terms of the recovery probability and the mean monotonically improve with the number of pilots.}
  \label{fig:NMSECDF}
\end{figure}

 For a sequence of $\Np=1024$ pilots, our joint estimation algorithm recovered the channel within acceptable limits with a probability greater than $0.95$, at a SNR of $0\, \mathrm{dB}$ using IID QPSK training. The number of outliers in this case is determined by the phase-transition region of PBiGAMP \cite{PBiGAMP}. It can be observed from \figref{fig:NMSECDF} that the probability of successful recovery monotonically increases as a function of the training length and quickly approaches $1$. 
To ignore the effect of outliers, the channel NMSE and CFO MSE results we report are averages over $95 \%$ of the realizations for $\Np=1024$. In practice, the failure probability can be lowered by increasing the training length or by designing a retransmission protocol that accounts for the failure. 
\begin{figure}[h]
\centering
\includegraphics[trim=1cm 6cm 2cm 7.5cm,clip=true,width=0.55 \textwidth]{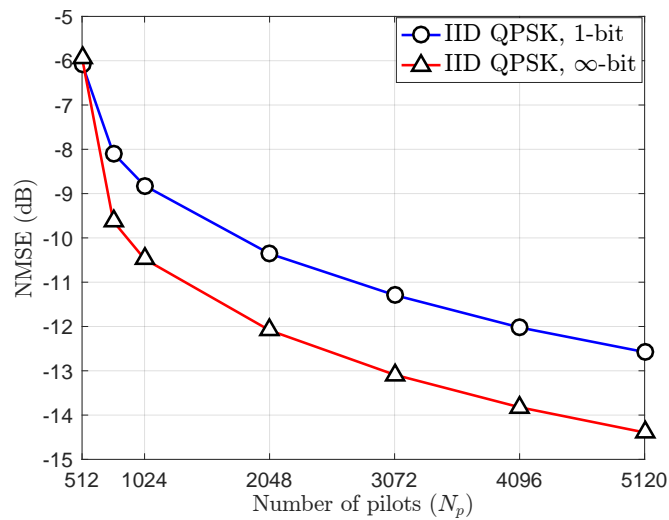}
  \caption{The NMSE of the channel estimate as a function of the training length, for an IID QPSK training at a SNR of $0$ dB. The NMSE monotonically decreases with the training length for one-bit and full resolution receivers. Due to the quantization noise in one-bit receivers, NMSE in the one-bit case is  higher than the full resolution one.}
  \label{fig:NMSEPIL}
  \vspace{-5pt}
\end{figure}
As seen in \figref{fig:NMSEPIL}, the NMSE monotonically decreases with the number of pilots. In practical wireless systems, the choice of the number of pilots is determined by the channel coherence time \cite{heathwicomm}.
\par
\begin{figure}[h]
\centering
\includegraphics[trim=1cm 6cm 2cm 7.5cm,clip=true,width=0.55 \textwidth]{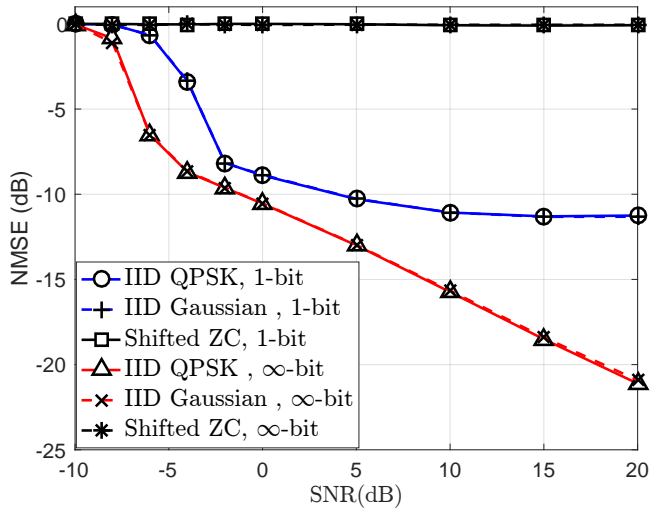}
  \caption{NMSE of the channel estimate obtained with one-bit and full resolution channel measurements for $\Np=1024$. Joint estimation is possible with IID Gaussian and IID QPSK training, but not with shifted ZC training due to the CFO propagation effect discussed in \secref{circulantTrn}. It can be observed that the  NMSE for the one-bit case saturates at high SNR due to quantization noise.}
  \label{fig:NMSESNR}
\vspace{-10pt}
\end{figure}

 In \figref{fig:NMSESNR}, we plot the NMSE as a function of the SNR for various training sequences. It can be observed that the reconstruction error is approximately the same for IID QPSK training and IID Gaussian training matrices. For the one-bit case, the NMSE saturates at high SNR because the recovery performance is limited by the quantization noise. As shown in \figref{fig:NMSEPIL}, the channel reconstruction error in one-bit receivers can be further decreased by using a higher number of pilots for the training. It can be noticed from \figref{fig:NMSESNR} that joint estimation with circulantly shifted ZC sequences proposed in \cite{mo2016channel} performs poorly. The failure due to such structured matrices can be attributed to the CFO propagation effect and confirms with our analysis in \secref{circulantTrn}. 
\subsection{MSE of the CFO estimate}
\begin{figure}[h]
\centering
\includegraphics[trim=1cm 6cm 2cm 7.25cm,clip=true,width=0.55 \textwidth]{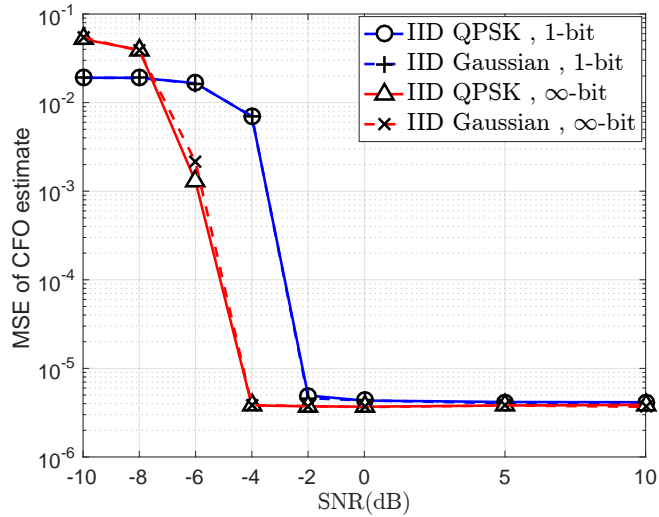}
  \caption{MSE of the CFO estimate as a function of SNR for a training length of $1024$. A maximally off-grid CFO of $\epsilon=45\pi/ 1024$ was chosen to evaluate the performance. At high SNR, the MSE saturates for both the one-bit case and the infinite resolution case due to phase noise \cite{salim2014channel}.}
  \label{fig:CFO_MSE_SNR}
\end{figure}

\begin{figure}[h]
\centering
\includegraphics[trim=1cm 6cm 2cm 7.25cm,clip=true,width=0.55 \textwidth]{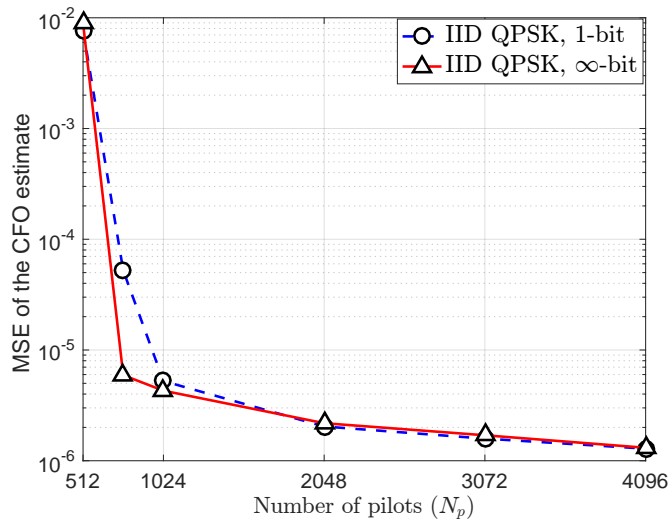}
  \caption{MSE of the CFO estimate as a function of the training length for a SNR of $0\, \mathrm{dB}$. Here, the CFO in the system was fixed to $\epsilon= 45 \pi / 1024$, and the phase noise standard deviation to $\beta= 0.067 \, \mathrm{rad}$. The MSE decreases with the number of pilots and the performance gap between the one-bit case and the full-resolution case is negligible. The sharp decrease in the CFO MSE can be attributed to the phase-transition effect of PBiGAMP \cite{PBiGAMP}. }
  \label{fig:CFO_MSE_PIL}
  \vspace{-10pt}
\end{figure}

The CFO in our algorithm is obtained using an Extended Kalman Filter \cite{EKF} on the inverse DFT of $\hat{\mathbf{b}}$. If $\hat{\epsilon}$ is the estimate of the CFO in the digital domain, the MSE of the CFO estimate is given by $\mathbb{E}\left[ \left( \epsilon -\hat{\epsilon}\right)^2\right]$. For a sequence of $1024$ pilots, the MSE of $\hat{\epsilon}$ is shown as a function of the SNR in \figref{fig:CFO_MSE_SNR}, for IID QPSK and IID Gaussian training matrices.
It can be noticed from \figref{fig:CFO_MSE_SNR} that the MSE of the CFO estimate saturates even for the full resolution case because the performance of the EKF is limited by the phase noise at high SNR \cite{salim2014channel}. Similar to the NMSE of the channel estimate, the MSE of $\hat{\epsilon}$ is expected to saturate at high SNR for one-bit receivers due to quantization noise. The CFO estimation error, however, is determined by the phase noise \cite{salim2014channel} because the CFO MSE for the one-bit case approaches that of the full resolution one. As expected, the CFO MSE decreases with the number of pilots and is shown in \figref{fig:CFO_MSE_PIL}.
\subsection{Performance invariance with the CFO}
\begin{figure}[h]
\centering
\includegraphics[trim=1cm 6cm 2cm 7.5cm,clip=true,width=0.55 \textwidth]{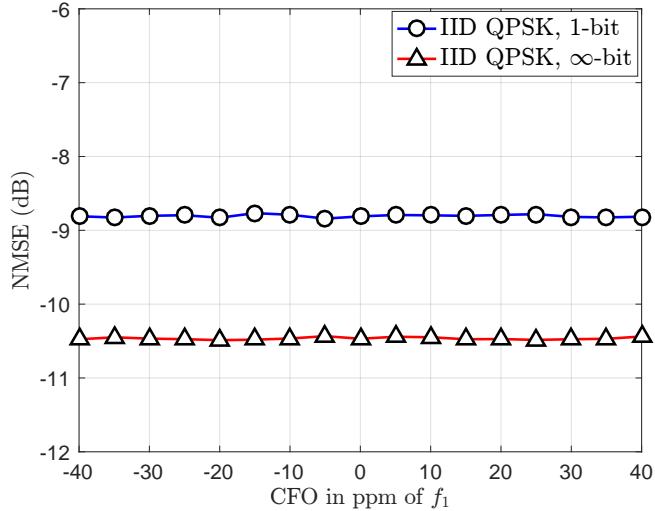}
  \caption{For a SNR of $0$ dB, the plot shows the invariance of the channel NMSE over the practical range of the CFO. The phase noise standard deviation was set to $0.067\, \mathrm{rad}$, and $1024$ pilots were used for the joint estimation. It can be observed from the plot that our joint estimation algorithm is robust to leakage effects that arise due to an off-grid CFO.}
\vspace{-10pt}
 \label{fig:NMSE_PPM}
\end{figure}

In this section, we show that the performance of our joint estimation approach is invariant to the CFO within practical limits. As an example, we evaluate our algorithm for different values of the CFO in the range $\left[-40 \, \mathrm{ppm}, 40 \, \mathrm{ppm}\right]$ of $f_1$. These limits were chosen according to the IEEE 802.11ad specifications. From \figref{fig:NMSE_PPM}, we see that the reconstruction errors in the channel is constant across the practical range of the CFO. The invariance arises due to the use of Bernoulli-Gaussian prior for the synchronization variables, that achieves robustness against off-grid leakage effects.
\section{Conclusions}
Most sparsity-aware channel estimation algorithms for mmWave systems assume perfect synchronization and perform poorly in the presence of such errors. In this work, we propose a message passing based algorithm that can leverage the sparsity of the wideband mmWave channel in addition to the compressibility of the phase error vector. We exploit the structure in the joint estimation problem to provide a low complexity implementation of a bilinear message passing algorithm. Unlike the existing methods that are specific to certain hardware architectures or use non-coherent techniques, our technique can be adapted to perform joint estimation with other mmWave architectures. 
\par Simulation results show that it is possible to perform joint CFO and channel estimation using IID QPSK and IID Gaussian training matrices. The CFO propagation effect proposed in this work shows that there is a trade-off between fast message passing using structured training and identifiability in joint estimation. In our future work, we will also consider frame synchronization while ensuring low complexity and scalability of our algorithm to different mmWave architectures. 
\small
\section*{Acknowledgment}
\normalsize
The authors would like to thank Jianhua Mo and Philip Schniter for helping them in getting started with the GAMP algorithm.
\vspace{-12pt}
\bibliographystyle{IEEEtran}
\bibliography{refs}
\end{document}